\documentclass[useAMS, subeqn, usenatbib]{mn2e}
\usepackage{epsf}
\usepackage{natbib}
\usepackage{graphicx}
\usepackage[section]{placeins}
\usepackage{epsfig}
\usepackage{hyperref}
\usepackage{comment}
\usepackage{multirow}
\usepackage{amsmath}
\usepackage{amssymb}
\usepackage{wasysym}
\usepackage{pdflscape}

\title[ULAS Proper Motion Catalogue]{A 1500 deg$^2$ Near Infrared Proper Motion Catalogue from the UKIDSS Large Area Survey}
\date{Month, 2013}
\author[L. Smith et al.]{Leigh Smith$^1$\thanks{E-mail: L.Smith10@herts.ac.uk}, P.W. Lucas$^1$, B. Burningham$^1$, H. R. A. Jones$^1$, R. L. Smart$^2$, 
\newauthor 
A. H. Andrei$^{1,3,4}$, S. Catal{\'a}n$^5$, D. Pinfield$^1$\\
$^1$ Centre for Astrophysics Research, Science and Technology Research Institute, University of Hertfordshire, Hatfield AL10 9AB, UK\\
$^2$ Istituto Nazionale di Astrofisica, Osservatorio Astrofisico di Torino, Strada Osservatorio 20, 10025 Pino Torinese, Italy \\
$^3$ Observat\'orio Nacional/MCTI, R. General Jos\'e Cristino 77, CEP 20921-400 Rio de Janeiro - RJ, Brazil.\\
$^4$ Observat\'orio do Valongo/UFRJ, Ladeira do Pedro Ant\^onio 43, CEP 20080-090 Rio de Janeiro - RJ, Brazil.\\
$^5$ Department of Physics, University of Warwick, Coventry, CV4 7AL
}

\begin{document}

\maketitle

\begin{abstract}
The UKIDSS Large Area Survey (LAS) began in 2005, with the start of the UKIDSS program as a 7 year effort to survey roughly 4000 deg$^2$ at high galactic latitudes in Y, J, H and K bands. The survey also included a significant quantity of 2-epoch J band observations, with an epoch baseline greater than 2 years to calculate proper motions. We present a near infrared proper motion catalogue for the 1500 deg$^2$ of the 2 epoch LAS data, which includes 135,625 stellar sources and a further 88,324 with ambiguous morphological classifications, all with motions detected above the 5$\sigma$ level. We developed a custom proper motion pipeline which we describe here. Our catalogue agrees well with the proper motion data supplied for a 300 deg$^2$ subset in the current WFCAM Science Archive (WSA) tenth data release (DR10) catalogue, and in various optical catalogues, but it benefits from a larger matching radius and hence a larger upper proper motion detection limit. We provide absolute proper motions, using LAS galaxies for the relative to absolute correction. By using local 2nd order polynomial transformations, as opposed to linear transformations in the WSA, we correct better for any local distortions in the focal plane, not including the radial distortion that is removed by the UKIDSS pipeline. We present the results of proper motion searches for new brown dwarfs and white dwarfs. We discuss 41 sources in the WSA DR10 overlap with our catalogue with proper motions $>$300 $mas~yr^{-1}$, several of which are new detections. We present 15 new candidate ultra-cool dwarf binary systems.
\end{abstract}


\section{Introduction}\label{intro}
	
	Stellar proper motion is the apparent angular movement of a star in a given time period. All stars have some component of motion (depending on the reference frame) due to their motion around the Galaxy and `gravitational kicks' they receive through interaction with other massive objects, usually molecular or atomic clouds. Motion perpendicular to a line between the star and the observer is the proper motion, which can be measured through careful observation of its position over two or more epochs, given sufficient time between observations dependant on instrument precision. 
	Given its relationship with distance and tangential velocity ($V_{tan} ~ \propto ~ d \cdot \mu$); a large proper motion is indicative of a fast moving and/or nearby source. For this reason many of the stars in the solar neighbourhood were first identified due to their large proper motion. 
	
	Major proper motion catalogues of the last half of the 20th century were developed using large scale surveys of Schmidt photographic plates often separated in time by many decades. 
	Large scale, deep, infrared sky surveys are very much a new thing as the size of infrared imaging arrays did not, until recently, permit them. The Two Micron All Sky Survey (2MASS; \citealt{skrutskie06}) and the Deep Near Infrared Survey of the Southern Sky (DENIS; \citealt{epchtein97}) are early examples of such surveys capitalising on recent improvements in infrared array technology. 2MASS and DENIS utilised 256$\times$256 pixel mercury cadmium telluride arrays. 2MASS used a pair of automated 1.3m telescopes, one in each hemisphere, and DENIS used a single 1m telescope at La Silla observatory in Chile. 
	Some proper motion catalogues have used near infrared data in conjunction with older optical catalogues to provide large epoch baselines, which improve the precision of the proper motion measurement, and also include accurate near infrared photometry (eg. PPMXL; \citealt{roeser10}, SIPS; \citealt{deacon05}). However, for a proper motion to be measured it must be detected in both surveys, meaning that very red objects which were not detected in the optical survey are missed.
	To overcome the problem of poor detectability of very red objects in such proper motion catalogues it is necessary to use infrared sky surveys alone. If we consider the use of 2MASS as the first epoch in a hypothetical near infrared only proper motion catalogue, then the current maximum epoch baseline of such a survey is 15 years. The astrometric accuracy of near infrared arrays is generally better than that of the Schmidt plates, which helps to offset the reduction in proper motion measurement precision due to shorter epoch baselines. Examples of current near infrared only proper motion surveys include a 2MASS only proper motion search \citep{kirkpatrick10} and a 2MASS - UKIRT Infrared Deep Sky Survey (UKIDSS) based proper motion search \citep{deacon09}. \citet{kirkpatrick10} identified 107 proper motion candidates that lack counterparts in Digitized Sky Survey B, R and I band images. Both examples have also identified a multitude of new nearby red objects (ultracool dwarfs), very few of which are detectable in current optical based surveys.
	
	Proper motion information is particularly useful when attempting to identify members of gravitationally bound systems. Their members serve as useful benchmark objects when one or more components of their systems have measurable attributes (eg. age, metallicity). Since members of such systems can be assumed to have formed from the same molecular cloud at a similar time, these attributes can also be inferred to belong to all members of a system \citep{pinfield06}. This is particularly useful in cases where it is difficult to constrain these attributes observationally, when dealing with ultracool dwarfs for example. Well characterised main sequence stars and white dwarfs make good companions for benchmark systems. Identification of a common proper motion and common distance is usually required to link multiple stars as single, gravitationally bound systems. 
	
	Ultracool Dwarfs (UCDs), generally regarded as spectral type M7 or later, are very low mass stars and brown dwarfs. They are chemically very interesting since their cool atmospheres allow dust and molecules to form.  A census of ultracool dwarfs is necessary to constrain the mass function at the substellar end, filling in the gap between giant planets and low mass stars (\citealt{burgasser04}, \citealt{pinfield06}, \citealt{kirkpatrick11b}). Ultracool dwarfs are usually selected photometrically in the infrared, often combined with optical photometry (\citealt{pinfield08b}, \citealt{burningham10}, \citealt{burgasser11}, \citealt{deacon11}, \citealt{dayjones11}), though spectroscopic confirmation is still necessary (\citealt{pinfield08b}, \citealt{dayjones11}, \citealt{kirkpatrick11a}). Proper motion is useful to discriminate between nearby ultracool dwarfs and background objects with similar colours such as high-redshift quasars and giant stars (\citealt{looper07}, \citealt{sheppard09}, \citealt{deacon11}).
	
	There are currently very few deep, wide field, near infrared proper motion surveys. This paper presents a new catalogue, which can be expected to reveal objects not detected in optical surveys while also providing kinematic data for known objects than can serve many scientific purposes, such as investigating the ages of T dwarfs \citep{smith13}.
\\	
\\This paper is organised as follows. In Section \ref{data} we describe the available data. In Section \ref{method} we describe our pipeline and construction of the catalogue which is available in the online data. A sample of the catalogue is available in the appendix of this paper. In Section \ref{sec:analysis} we determine the accuracy and reliability of the catalogue and discuss known limitations. In Section \ref{sec:results} we outline searches undertaken for objects of interest within the catalogue. In Section \ref{newbenchmarkcands} we reveal interesting sources identified during searches for multiple systems. In Section \ref{conclusion} we draw conclusions.

\section{Data}\label{data}

The United Kingdom Infrared Deep Sky Survey (UKIDSS, \citealt{warren02}, \citealt{lawrence07}) project began in 2005, and was a 7 year effort to survey approximately 7000 deg$^2$ using the 3.8m infrared-dedicated United Kingdom Infra-Red Telescope (UKIRT), situated at the summit of Mauna Kea, Hawaii, and the Wide Field CAMera (WFCAM, \citealt{casali01}, \citealt{casali07}).

The WFCAM consists of four 2048${\times}$2048 pixel arrays, which combined with UKIRT optics give a total viewing area of 0.21 deg$^2$ (0.4" per pixel, \citealt{casali07}). During observation the arrays were micro-stepped for the UKIDSS LAS J band, four individual exposures are taken, each with a 0.5 pixel offset in x and/or y from the first and recombined during the Cambridge Astronomy Survey Unit (CASU) pipeline using a process called interleaving \citep{vick04}. Interleaving is performed using a process called \textsl{dribbling}, which corrects point spread function (PSF) mismatches caused by changes in the observing conditions between exposures, which can lead to a `spiky' PSF \footnote{CASU, \href{http://casu.ast.cam.ac.uk/surveys-projects/wfcam/technical/interleaving}{http://casu.ast.cam.ac.uk/surveys-projects/wfcam/technical\\/interleaving}}. This process of over-sampling improves the resolution of the WFCAM images to the limit of the seeing. The WFCAM photometric system is described in detail in \citet{hewett06}. After the CASU pipeline the data are then transferred to the WFCAM Science Archive (WSA, \citealt{hambly08}) for further processing and to make the data available for the community.

The LAS covers 4028 deg$^2$ in YJHK passbands to an approximate 5$\sigma$ depth of 19.6 in J and is complemented in the ugriz optical passbands by the Sloan Digital Sky Survey (SDSS). The LAS included a second epoch of observations in the J passband to calculate proper motions and investigate stellar variability. In the final months of the UKIDSS program great effort was made to observe as much of the first epoch coverage as possible at second epoch. The final second epoch coverage is around 1500 deg$^2$. 

UKIDSS LAS multi-frame catalogues based on J band images taken during the period 2005 May 15$^{th}$ until 2012 May 20$^{th}$ were obtained from the WFCAM Science Archive \citep{hambly08} and paired using the telescope pointing coordinates to identify coincident multi-frames. In many cases several repeats of each pointing had been obtained over a relatively short period of time (typically days, weeks or months). This reflects the fact that multi-frames may be rejected as part of the at-the-telescope survey quality control, and thus queued for repeats, but are still processed and committed to the archive. To ensure that the best quality frames were used for our first and second epochs at each pointing and avoid the use of deprecated frames, we only accepted pairs of multi-frames where both multi-frames represented the latest date amongst data taken at each epoch. This resulted in typical epoch baselines between multi-frames of between 1.8 and 7 years. 

\begin{figure}
\begin{center}
\begin{tabular}{c}
\epsfig{file=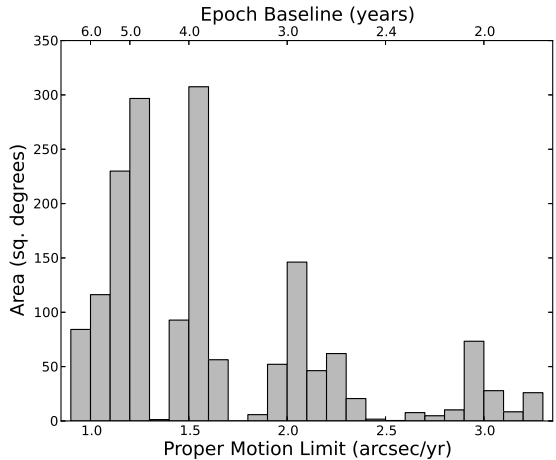,width=1\linewidth,clip=}
\end{tabular}
\caption{Plot showing the area distribution of epoch baselines and proper motion limits of the catalogue. Note that the total area represented by this plot is 1678 deg$^2$, this is simply the number of frames multiplied by the angular area of a frame, the catalogue is nearer to 1500 deg$^2$ after removal of the duplicated regions of frame overlaps.}
\label{pmlimit}
\end{center}
\end{figure}

We constructed two epoch catalogues for each pointing by matching sources within the pairs of multi-frames using the Starlink Tables Infrastructure Library Tool Set (STILTS; \citealt{taylor06}). We required pairs of sources to be uniquely paired to their closest match within 6", and we required the J band magnitudes for the two epochs to agree within 0.5 magnitudes, to minimise mismatches. 
Given the minimum epoch baseline of 1.8 years, the hard proper motion limit of the catalogue is therefore 3."3$~yr^{-1}$ though the catalogue is built from frame sets with a range of epoch baselines, giving a range of proper motion limits. Figure \ref{pmlimit} shows the area distribution of the epoch baselines and the corresponding proper motion limits. Note that we performed an initial rejection of the few input sources brighter than 12$^{th}$ or fainter than 20$^{th}$ magnitude in the J band (see Section \ref{catcaveats}).

\section{Method}\label{method}

\subsubsection*{Overview}
The method involves selecting a sample of good reference stars based on a variety of astrometric and photometric cuts. We then fit the motion
between the two epochs using a second order polynomial either locally or across the whole array, depending on the local source density and proximity to the edge of the array. Motions of most sources are calculated using a unique local fit to stars well distributed around them. We use local transformations in preference as they produce more accurate results (see Figure \ref{nf1utferr}).
	
\subsubsection*{Definitions}	

In this paper we adopt the following terms, consistent with those used by the WSA.
\begin{description}
\item Frame - An image or catalogue data from one of the four WFCAM arrays.
\item Frame set - A set of frames covering the same area and multiple bandpass and/or epochs.
\item Multi-frame - A set of four frames comprising one whole WFCAM footprint in one bandpass and epoch (exclusive of the guider chip).
\end{description}

For the purpose of this description we adopt the following terms:
\begin{description}
\item Global (fit/transform) - The operation was performed using all relevant data in one frame.
\item Local (fit/transform) - The operation was performed using a limited area of one frame.
\item Target (source/frame) - Where an operation is performed on each source/frame individually, we refer to an example as the {\it target} source/frame.
\item J1 and J2 - Refer to the first and second epoch J band images respectively.
\end{description}

\subsection{Reference Star Selection}\label{cppselection}

A preliminary pool of astrometric reference sources was created as a subset of the input catalogue, containing sources meeting the following criteria:
\begin{description}
\item Classified as stellar at J1 \& J2;
\item J1 \& J2 between 16 \& 19.6;
\item J1 \& J2 magnitude error $<$ 0.1; and
\item J1 \& J2 ellipticity $<$ 0.3
\end{description}
We rejected frames containing fewer than 20 reference sources. The minimum requirement for a second order polynomial fit is 6 but we adopted 20 to ensure the data were well fitted across the frame. In practice we rejected 217 frame sets (0.65\%), losing 46,097 sources (0.26\%) at this stage.

On a frame by frame basis we fit the second epoch array (x/y) positions of the reference sources to their first epoch array positions using a second order polynomial fit and the \textsc{CP2TFORM} function in \textsc{MATLAB} to produce a preliminary global transform. We applied the inverse of this coordinate transformation(\textsc{MATLAB} does not allow a forward transformation for a second order polynomial) to map the first epoch reference source positions on to the second epoch positions and subtracted these from their second epoch positions to produce preliminary residuals. We measured the uncertainty on the preliminary residuals by calculating the RMS residual to the fit of all reference sources in each frame and added these in quadrature to their centroid errors.

We rejected all reference stars with significant preliminary residuals ($>$1$\sigma$) usually indicating motion. We then discarded all preliminary positions and motions and performed a further rejection of frames failing the minimum 20 reference stars cut. A further 144 frame sets (42,415 sources) were rejected at this stage taking the frame and source counts to 33,038 and 17,122,488 respectively.

Note that we use array coordinates to calculate the motions since the astrometric fit of LAS frames is performed by CASU using the positions of 2MASS sources, which were observed near epoch 2000. The quality of these fits has degraded over time due to the motions of the 2MASS sources used.

\subsection{Second Epoch Position Correction}\label{pmcalculation}

Final residuals are calculated on a source by source basis. We select all reference stars (with the exception of the target source) in the same frame as the target source as a temporary pool of reference stars. We calculate a global transform by fitting the first epoch reference star array positions to the second epoch reference star array positions using a second order polynomial as before, and apply the inverse coordinate transformation to the second epoch target position to map them on to the first epoch array coordinate system. We then calculate the RMS residual to the fit of the reference sources and add it in quadrature to the centroid error of the target at the second epoch to calculate the uncertainty on the transformed position.

Another second order polynomial fit was then calculated and applied in the same manner but using only reference stars local to the target. We selected all reference stars within a radius sufficient to ensure that there were at least 3 in every attached circle quadrant. This radius was rounded up to the nearest 20" and we impose a minimum radius of 1'.

The use of this method ensured that there were at least 12 reference stars used to calculate each fit and crucially that the reference stars were well distributed about the target source. If any quadrant contained fewer than 3 reference stars then a local polynomial was not calculated and we default to using the global polynomial to calculate a final proper motion. This was always the case for sources at the edge of frames. A `true' value in the \textsl{local} column of the catalogue indicates that a source has a proper motion calculated using a local transform. We applied the local polynomial to the target source's second epoch position to map it on to the first epoch array coordinate system. We then follow the same uncertainty calculation method as before.

\begin{figure}
\begin{center}
\begin{tabular}{c}
\epsfig{file=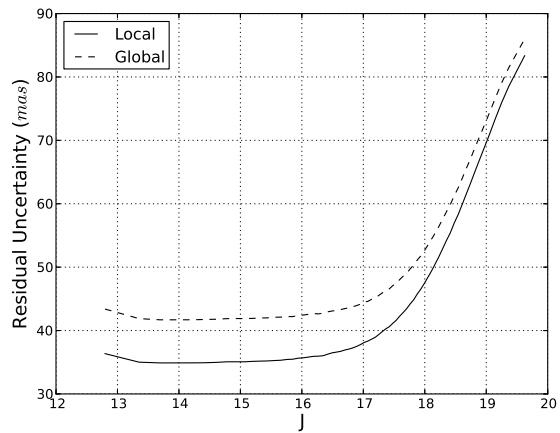,width=1\linewidth,clip=}
\end{tabular}
\caption{We selected sources meeting the criteria in Section \ref{sec:laslascomp} with measured local and global residuals ($3.5$ million sources total) and split them into 70 equal sized J magnitude bins. Note that the exact width in J magnitudes of each bin was allowed to be different, giving greater resolution where the source density allowed but maintaining accuracy at the extreme ends. The mean local and global residual uncertainties in each bin are shown. The local residual uncertainties are consistently lower than the global ones.}
\label{nf1utferr}
\end{center}
\end{figure}

To calculate proper motion we used residuals calculated from the local transforms in preference to the global ones. We justify this preference by looking at the uncertainties on the total residuals for the two samples (see Figure \ref{nf1utferr}), where the local transform produces smaller average uncertainties on the residuals than the global transform.

\subsection{Conversion to Equatorial Coordinate System}\label{eqcoordsystem}

In order to transform the array coordinate positions on to the tangent plane to the equatorial system, the first epoch $\alpha \slash \delta$ positions underwent a tangent plane projection conversion about the centre of the frame, producing $\xi \slash \eta$ positions (\ref{xiequation}, and \ref{etaequation}).

\begin{subequations}
\begin{align}
\xi &= \frac{\cos \delta \sin(\alpha - \bar{\alpha})}{\sin \delta \sin \bar{\delta} + \cos \delta \cos \bar{\delta} \cos(\alpha - \bar{\alpha})} \label{xiequation}\\
\eta &= \frac{\sin \delta \cos \bar{\delta} - \cos \delta \sin \bar{\delta} \cos(\alpha - \bar{\alpha})}{\sin \delta \sin \bar{\delta} + \cos \delta \cos \bar{\delta} \cos(\alpha - \bar{\alpha})} \label{etaequation}
\end{align}
\text{where $\bar{\alpha}, \bar{\delta}$ are the centre points of the frame in the $\alpha, \delta$ dimensions}\\
\end{subequations}

We then fit the first epoch $\xi$/$\eta$ positions of all sources in the frame to their corresponding array positions using a third order polynomial and then applied its inverse to the first and second epoch array positions (both now in the first epoch array coordinate system) to transform them on to first epoch tangent plane. This was simpler than applying the $\alpha$, $\delta$ information in the fits headers to the second epoch data and has a precision better than 1 $mas$. Creation of a polynomial on which differentiation can be performed from the transformation matrix created in this process is not trivial. It is therefore very difficult to formally propagate the uncertainties through the transformation. Instead we transformed the array position 1$\sigma$ error box, the uncertainty being half the difference between these boundaries after the transformation was applied. Calculation of each source's proper motion was then a matter of subtracting the first epoch tangent plane positions from the second epoch tangent plane positions and dividing through by the epoch baseline. The uncertainty on the proper motion is the first and second epoch positional uncertainties added in quadrature and divided by the epoch baseline.

\subsection{Relative to Absolute Proper Motion Correction}\label{rel2abscorr}

\begin{figure*}
\begin{center}
\begin{tabular}{c}
\epsfig{file=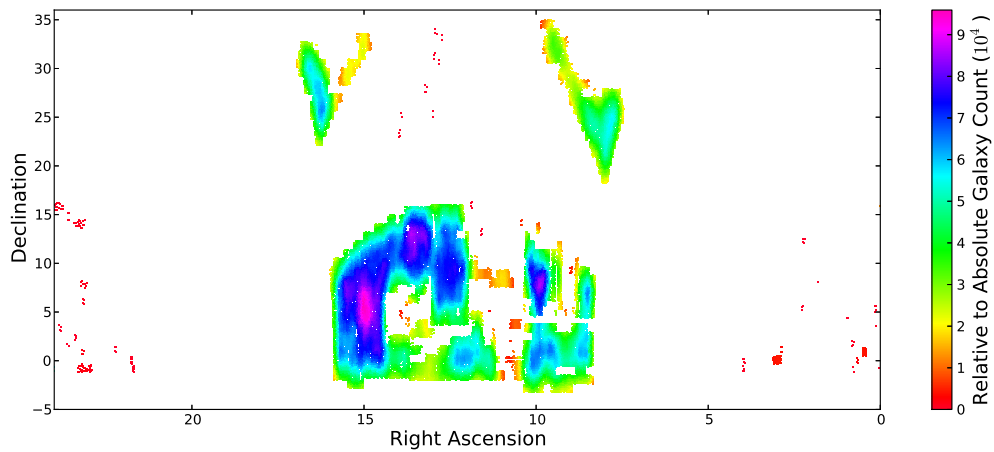,width=1\linewidth}
\end{tabular}
\caption{The distribution of the number of galaxies used to calculate the relative to absolute correction of each frame. Relatively few are used in isolated frames; the lowest value is 5. In frames central to the larger fields values can be as high as 95,000.}
\label{r2acnts}
\end{center}
\end{figure*}

Until this stage proper motions were relative to the mean motion of the reference sources used for the fit. These were stellar sources which all have a component of proper motion due to galactic rotation and solar motion. We remove this component of proper motion and convert the relative proper motions to absolute ones, defined by selected extragalactic sources. We calculated the median relative proper motion of sources meeting the following criteria:
\begin{description}
\item Classified as a galaxy in J1 \& J2;
\item J1 \& J2 between 12 \& 19.6;
\item J1 \& J2 magnitude error $<$ 0.2; and
\item Total relative proper motion error $<$ 30 $mas~yr^{-1}$
\end{description}
We used sources in the target frame and those from surrounding frames within three degrees. Their median motions were then subtracted from the relative proper motions of all sources in the target frame. We find that using extragalactic sources only in the same frame or using the mean relative motion for all sources within three degrees introduces significant local scatter in the correction vectors due to inaccuracies in the centroids of extended objects. Figure \ref{r2acnts} shows how the number of galaxies used varies with sky position. No correction is greater than 10 $mas~yr^{-1}$ in $\mu_{\alpha} \cos \delta$ or 12 $mas~yr^{-1}$ in $\mu_{\delta}$. This is typically less than the uncertainties on the motions. Ideally quasars located in the same frame would be used to calculate the correction, however we would require a sample of confirmed quasars with several well distributed in each frame. The standard error on the median of the relative proper motion uncertainty of the selected galaxies was then added in quadrature to the uncertainties of the relative proper motions of all sources in the target frame to calculate the uncertainties on the absolute proper motions. The median contribution of the relative to absolute proper motion correction to the absolute proper motion uncertainty is 0.016 $mas~yr^{-1}$ in both dimensions.

\subsection{Duplicate Source Removal}\label{dupremoval}

\begin{figure}
\begin{center}
\begin{tabular}{c}
\epsfig{file=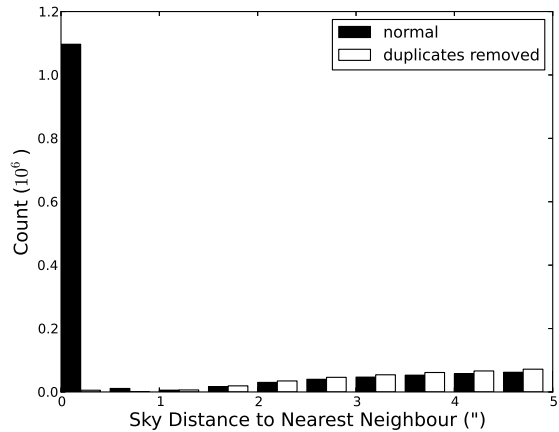,width=1\linewidth,clip=}
\end{tabular}
\caption{The distribution of distances between sources (within 10") before and after application of our duplicate removal method. The huge peak at very low separations and the fact that overwhelming majority of these pairs are in different frames is indicative of duplicate sources in frame overlap regions. After removing these duplicates using the method in section \ref{dupremoval} the peak has been almost entirely removed. The remaining sources are close pairs in the same frame set therefore likely to be genuine.}
\label{wwodups}
\end{center}
\end{figure}

The catalogue contained duplicates of sources in regions of overlapping frames. We matched internally for groups of sources with separations less than 1" using the Tool for OPerations on Catalogues And Tables (TOPCAT; \citealt{taylor05}), finding 1,614,695 initial groups containing a total of 3,380,822 sources. We found that 99.94\% of groups with separations of 0."5 or less contained sources from different frames.
We made the assumption that since the overlap of the frames is typically $\sim$24" it is unlikely that genuine neighbouring sources would be present only on different frames. Instead, both components of a genuine group would be duplicated. Using this assumption we remove all but the source with the lowest uncertainty on the total proper motion from groups containing sources from different frames (see Figure \ref{wwodups}). This removed all but the most well measured source from each set of duplicates, a reduction in catalogue size by 10.6\%.

\subsection{YHK Retrieval and Bad Data Removal}

We matched LAS DR10 first epoch J band equatorial positions and magnitudes retrieved from the WSA lasDetection tables to our catalogue, giving us WSA assigned source IDs and hence a method to accurately match to their source table and retrieve the data contained within. We retrieved Y, H, and K magnitudes and their associated uncertainties as well as first and second (where available) epoch J band post processing error bits (ppErrBits\footnotemark[1]) information. ppErrBits is a useful indicator of the quality of each detection, larger values are indicative of more severe detection quality issues. We removed from the catalogue all sources with ppErrBits values of 256 or greater which would correspond to saturation or electronic crosstalk \citep{dye06} or poor flat field region, etc.

\subsection{Bad Pixel Sources}

\begin{figure}
\begin{center}
\begin{tabular}{c}
\epsfig{file=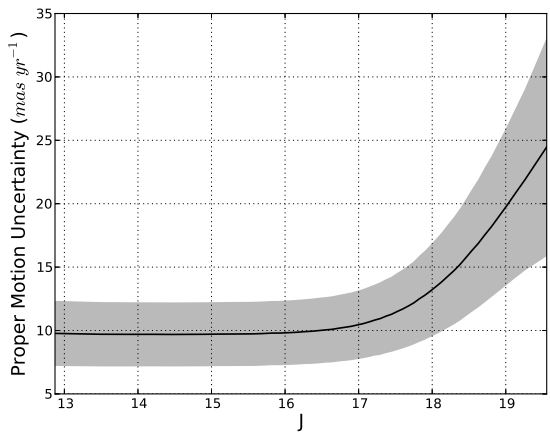,width=1\linewidth,clip=}
\end{tabular}
\caption{We selected sources meeting the same criteria as in Section \ref{sec:laslascomp} ($5.4$ million sources) and split them into 53 J magnitude bins each containing approximately 100,000 sources but having a variable width. The mean uncertainty on $\mu_{total}$ in each bin is plotted. The shaded section shows the region bound by 1 standard deviation.}
\label{pmerrs}
\end{center}
\end{figure}

Approximately 20\% of catalogue sources have a `$-7$' (bad pixel within 2" aperture) classification at either epoch. We find this has a negative effect on the precision of the astrometry, as one might expect. The median total proper motion for this selection is 50\% larger than that of the rest of the catalogue, whereas the mean uncertainty is only 25\% larger. We expect the source with the median total proper motion will in reality have a negligible motion and as such the mean uncertainty on the value should be of a similar magnitude. Although the proper motion uncertainties on sources with a bad pixel classification at either epoch was already slightly higher than normal sources (by this factor of about 1.25), we inflated their proper motion uncertainties by a factor of 1.2 to mirror the relative increase in median proper motion by this amount. The distribution of the final uncertainties on absolute proper motions is shown in Figure \ref{pmerrs}.

The presence of a ``$-7$'' in the classification column means that a genuine classification ($-1$/$1$, stellar/extragalactic, etc) is unavailable. To compensate for this we include the WSA merged class attribute\footnotemark[1] where available. Merged Class is a combination of classifications in all available bands and epochs of UKIDSS DR10 using Bayesian classification rules. 

We note that since the proper motions of these objects are less reliable they are not used as reference sources at any stage of the pipeline.

\footnotetext[1]{see \href{http://surveys.roe.ac.uk/wsa/ppErrBits.html}{http://surveys.roe.ac.uk/wsa/ppErrBits.html}}

\section{Analysis of Results}\label{sec:analysis}

While we produce results for all LAS sources here we publish only those with absolute proper motions detected at the 5$\sigma$ level and above, with a morphological classification indicating a likely stellar nature. We include sources classified as stellar (class $=$ -1) or probably stellar (class $=$ -2) at one or more epochs and exclude sources classed as noise (class $=$ 0) at either epoch. We find 135,625 sources classified as stellar in both J band detections, and a further 88,324 sources with ambiguous morphological classifications. This produces a catalogue of 223,949 sources in the 1500 deg$^2$ area shown in Figure \ref{r2acnts}. Note that ellipticity and morphological classification trace genuine high proper motion detections very well at J$\le$19 (see Section \ref{reliability}). However, in the interests of not rejecting large numbers of potentially genuine sources we impose no restriction on ellipticity. We recommend that users employ cuts on ellipticity and morphological classification if a very reliable high proper motion sample is sought. The lowest uncertainties for the brightest and faintest sources are 4 and 12 $mas~yr^{-1}$ respectively, corresponding to the longest epoch baselines. The 5$\sigma$ lower limit on absolute proper motion significance therefore corresponds to minimum proper motions of 20 to 60 $mas~yr^{-1}$ for bright and faint sources respectively. A sample of the catalogue is presented in the appendix and the full table is available in the online data.

We scrutinised 1/5th of the results, approximately 300 deg$^2$. This area corresponds to the overlap with second epoch J coverage of UKIDSS DR10.

\subsection{Comparison to WSA Proper Motions}\label{sec:laslascomp}

\begin{figure}
\begin{center}
\begin{tabular}{c}
\epsfig{file=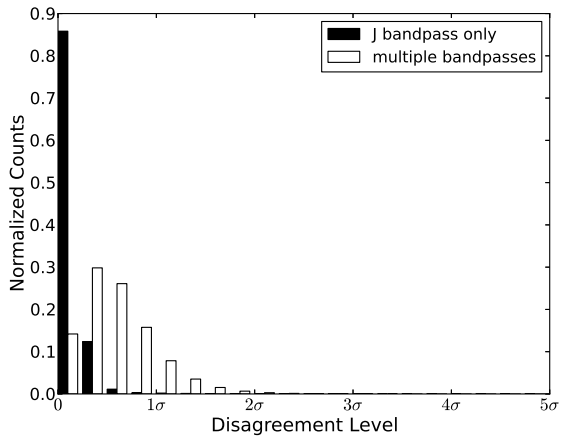,width=1\linewidth,clip=}
\end{tabular}
\caption{Histogram showing the distribution of disagreement between proper motions of stellar sources from our pipeline and those from the WSA. The catalogues agree very well where the WSA has used only J bandpass data for their proper motion, and less well where they have use multiple bandpasses. Pearson product-moment correlation coefficients are 0.99 and 0.80 for J only and multiple bandpass total proper motions respectively.}
\label{wsacomparison}
\end{center}
\end{figure}

With the WFCAM Science Archive's 9th release of LAS data came proper motions \citep{collins12} to which we have compared our results (Figure \ref{wsacomparison}). The WSA proper motions are not absolute, so here we compare using the relative proper motions calculated by our pipeline. The WSA uses a linear transform in the tangent plane across the whole frame which we have shown in Section \ref{pmcalculation} to be less accurate. It effectively assumes that there are no non-linear distortions in the focal plane apart from the known 3rd order radial distortion that is removed by the UKIDSS pipeline as part of the astrometric solution for each WFCAM array.

We created and calculated proper motions for a new input data set containing the most recent WSA DR10 data from the LAS detection table. Matching the two catalogues using the unique source IDs assigned by the WSA and maintained throughout our proper motion pipeline ensures there are no mismatches.

We select sources with no post processing error flags, low ellipticity and classified as stellar in both J band images as an appropriate group of sources for comparison, a total of 1.6 million sources.

The proper motion measurements are fairly consistent between the catalogues with Pearson product-moment correlation coefficients of 0.80 and 0.82 in $\mu_{\alpha} \cos \delta$ and $\mu_{\delta}$ respectively and $86\%$ and $99\%$ of proper motions matching within their $1\sigma$ and $2\sigma$ uncertainties respectively. The WSA proper motions are obtained using all available LAS detections in the YJHK passbands. The WSA assumes that chromatic dispersion is minimal, and hence no effort is made to correct for the effects of this. We note that where the WSA results used multi-band frames to calculate a proper motion our values differ slightly more, with Pearson's r coefficients of 0.79 and 0.82 in $\mu_{\alpha} \cos \delta$ and $\mu_{\delta}$ respectively. As one might expect, for the few sources with only J band images the proper motions agree very well, with Pearson's r coefficients of 0.99 in both $\mu_{\alpha} \cos \delta$ and $\mu_{\delta}$.

\subsection{Comparison to LSPM Catalogue}\label{sec:lspmcomp}

The LSPM catalogue \citep{lepine05} utilises the SUPERBLINK software \citep{lepine02} to normalise the differences between pairs of sub-frames from the POSS-I \citep{abell59} and POSS-II \citep{reid91} plates (usually involving a degradation in the quality of the POSS-II plate to match the POSS-I plate quality), then subtraction of one from the other to produce a residual image which maps the first and second epoch positions of sources with high proper motion. The catalogue benefits from the fact that all high proper motion sources identified by the SUPERBLINK software were manually blinked to remove any erroneous high proper motion sources, the LSPM catalogue has a minimal false detection contamination as a result. The LSPM catalogue also includes data from the TYCHO-2 \citep{hog00} catalogue and the All-Sky Compiled Catalogue \citep{kharchenko01}.

\begin{figure}
\begin{center}
\begin{tabular}{c}
\epsfig{file=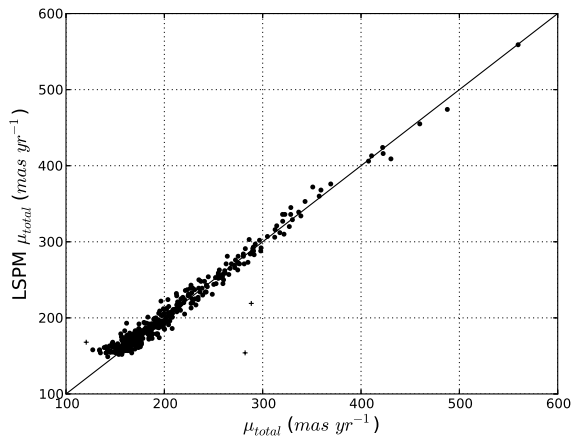,width=1\linewidth,clip=}
\end{tabular}
\caption{LSPM total proper motions (vertical axis) versus those calculated by our pipeline (horizontal axis) for the 381 matches between the two catalogues. The crosses are LSPM J1644+3203, LSPM J1625+2519, and LSPM J1609+2457 for which the total proper motions differ greater than 4$\sigma$. The data are nevertheless well correlated; the Pearson product-moment correlation coefficient is 0.980.}
\label{lspmcomparison}
\end{center}
\end{figure}

We matched the LSPM-North catalogue to our LAS proper motion catalogue using a 3" matching radius and a 0.5 magnitude J band discrepancy tolerance. We find 381 matches and compare LSPM and our LAS proper motions, see Figure \ref{lspmcomparison}. The majority of LSPM proper motions given are derived using the author's SUPERBLINK software, there is one Tycho-2 proper motion and four from `other' sources, these five proper motions agree well with those from our LAS catalogue. We found proper motions from both catalogues agreed within their $1\sigma$ uncertainties for $79\%$ of sources, this rises to $98\%$ agreement at $2\sigma$. The LSPM proper motion uncertainties were taken as 8mas/yr \citep{lepine05}. The proper motions are also well correlated, with Pearson product-moment correlation coefficients of 0.994, 0.979 and 0.980 ($\mu_{\alpha} \cos \delta$, $\mu_{\delta}$ and $\mu_{total}$ respectively). Statistically, from a sample of 381 sources we do not expect any to have proper motions with a disagreement greater than $4\sigma$, we find 3: LSPM J1644+3203, $4.43\sigma$; LSPM J1625+2519, $4.81\sigma$; and LSPM J1609+2457, $27.41\sigma$.

\subsubsection*{LSPM J1644+3203}
In the J1 image the high proper motion source is overlapping another source to the north with a separation of 1."3. This is probably causing the centroid on the object at J1 to be skewed north causing the observed larger proper motion in declination. The proper motion in right ascension agrees comfortably. LSPM J1644+3203 is NLTT 43473 (see Section \ref{sec:rnlttcomp}) which has a proper motion in agreement with the LSPM catalogue. 

\subsubsection*{LSPM J1625+2519}
On inspection of the 2 epochs of UKIDSS LAS J band images the source is separated by 1."7 from another source, which was unresolved in the photographic data. Plotting the positions of the centroids at both epochs shows that at the second epoch the centroids are well fitted to both sources. The first epoch image quality is slightly lower which caused the fainter target to go undetected and the centroid for LSPM J1625+2519 to be skewed towards it, altering the measured proper motion.  Interestingly the source which is overlapping LSPM J1625+2519 appears to share a common proper motion with it.

\subsubsection*{LSPM J1609+2457}
While blinking the 2 epochs of UKIDSS LAS images the source does appear at first glance to exhibit a proper motion consistent with our value, we note that the quality of the first epoch J band image is poor. Blinking of the POSS I and II images reveals a motion consistent with the value given in the LSPM catalogue. No other source with a proper motion consistent with LSPM J1609+2457 is found in our catalogue within 1' of its given location. The cause of this erroneous proper motion measurement is likely the poor first epoch UKIDSS J band image and resultant centroid fit. Since the LAS proper motion is likely to be the incorrect proper motion measurement we have included this source in our comparison.

\subsubsection*{}
We attribute the presence of the poor quality images mentioned above to our use of data that have not yet been through the UKIDSS quality control procedures that take place prior to a formal SQL data release.  We note that this has probably been the cause of two of the erroneous proper motions from this sample of 381.

\subsubsection*{}
In an effort to gauge the completeness of the catalogue we identified LSPM sources within the UKIDSS DR10 area and with 2MASS J magnitude $>$ 12.5. The J magnitude cut includes null values and allows for a half a magnitude discrepancy between the UKIDSS and 2MASS J band magnitudes, this is necessary to accomodate recovery of UKIDSS objects measured up to half a magnitude brighter than in 2MASS, which would otherwise appear unrecovered due to our 12th magnitude bright limit. We identify 379 LSPM sources that should be present within this catalogue, of which we recover 375 with proper motions that agree within 4$\sigma$. A further three sources have discrepant proper motions, these are LSPM J1644+3203, LSPM J1625+2519, and LSPM J1609+2457 (see above). The final unrecovered source is LSPM J0829+2539/LHS 2015. LHS 2015 is a previously unresolved common proper motion pair originally classified by \citet{reid05} as a DQ white dwarf. The pair are unresolved in the first epoch J band image and are consequentially more than half a magnitude brighter than the resolved magnitudes at the second epoch. This caused the pair to fail this quality control cut at the epoch matching stage. If we consider sources with discrepant proper motions as unrecovered then we have an LSPM source recovery rate of 98.9\%, otherwise the recovery rate is 99.7\%.

\begin{figure}
\begin{center}
\begin{tabular}{c}
\epsfig{file=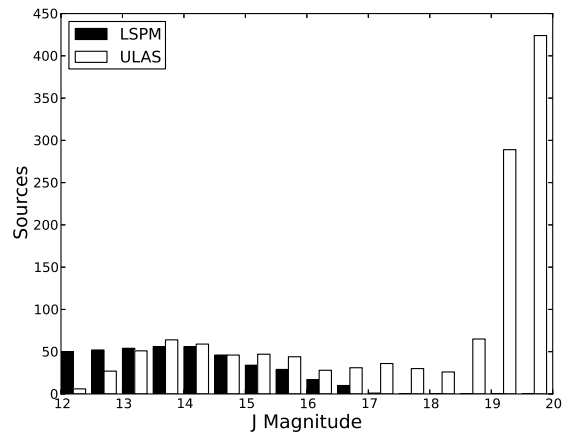,width=1\linewidth,clip=}
\end{tabular}
\caption{A comparison between the number of LSPM high proper motion stars and those from this catalogue after application of ellipticity and morphological classification cuts (see text). Note that the false positive rate of the ULAS high proper motion detections increases sharply at J$\sim$19. Our catalogue is more complete fainter than J$=$13.5.}
\label{lspmulas}
\end{center}
\end{figure}

Figure \ref{lspmulas} compares the number of high proper motion ($>$150 $mas~yr^{-1}$) sources fainter than J$=$12 as a function of J magnitude in the UKIDSS DR10 area from our catalogue and the LSPM. We require sources in our catalogue to have ellipticity $<$ 0.3 and be classified as stellar at both epochs. This requirement means we can infer an approximate false positive rate from Figure \ref{reliability_scatter}, at the likely expense of some genuine detections. The LSPM catalogue is more complete at the bright end, where our catalogue suffers from near saturation. We begin to find more high proper motion sources than the LSPM catalogue at about J$=$13.5. The false positive rate of ULAS sources increases sharply at around J$=$19 (see Figure \ref{reliability_scatter}) which must be taken into account, and that 25 LSPM sources (6\%) have null J band magnitudes and therefore could not be included in the comparison. The decline in source counts in our catalogue at J$\ge$13.5 we believe is due to the increasing average distance, and hence smaller average proper motion of these relatively faint stars.

\subsection{Comparison to Revised NLTT Catalogue}\label{sec:rnlttcomp}

The Luyten Half Second (LHS) Catalogue contains stars with proper motions exceeding 0."5 annually \citep{luyten79a}. The LHS catalogue contains positions, proper motions and optical magnitudes for 4,470 stars with proper motions greater than 239 $mas~yr^{-1}$ (note that a small number of sources were included, in spite of the 0.5 "$~yr^{-1}$ lower limit). The catalogue includes data compiled from other proper motion searches and 804 hand/machine-blinked Palomar Sky Survey fields. The LHS catalogue was revised by \citet{bakos02}. 4,323 of the original 4,470 high proper motion sources were recovered and their positions and proper motions were refined. 

The NLTT (New Luyten catalogue of stars with proper motions larger than Two Tenths of an arcsecond, \citealt{luyten79b}) catalogue is an extension of the LHS catalogue down to proper motions of 40 $mas~yr^{-1}$ for 58,845 sources. A minority (152) have proper motion less than 180 $mas~yr^{-1}$ however. The NLTT catalogue was revised and refined by \citet{salim03}, giving improved positions and proper motions for sources present in both the original POSS I frames and the second 2MASS data release.

\begin{figure}
\begin{center}
\begin{tabular}{c}
\epsfig{file=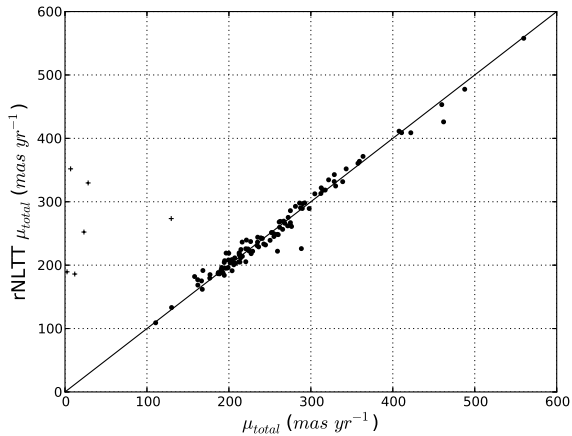,width=1\linewidth,clip=}
\end{tabular}
\caption{Comparison between proper motions from the revised NLTT catalogue (vertical axis) and those calculated by our pipeline (horizontal axis) for the 109 reliable matches between the catalogues. The 6 sources represented by crosses are those with proper motions differences greater than 2$\sigma$ listed in section \ref{sec:rnlttcomp}, these were removed from the comparison since they were found to have spurious rNLTT proper motion values. The Pearson product-moment correlation coefficient for the remaining data is 0.988.}
\label{rnlttcomparison}
\end{center}
\end{figure}

We matched the Revised NLTT catalogue (rNLTT, \citealt{salim03}) to our LAS proper motion catalogue using the same matching criteria as for the LSPM comparison (section \ref{sec:lspmcomp}), this time finding 115 initial matches, see Figure \ref{rnlttcomparison}. We find proper motions from both catalogues agreed within their $1\sigma$ uncertainties for $70\%$ of sources, rising to $94\%$ agreement at $2\sigma$. The remaining 7 sources have proper motion disagreements of greater than their $4\sigma$ uncertainties: NLTT 43473, $4.73\sigma$; NLTT 22010, $6.22\sigma$; NLTT 21214, $9.36\sigma$; NLTT 20123, $10.14\sigma$; NLTT 18649, $21.01\sigma$; NLTT 18692, $21.52\sigma$; and NLTT 19021 $26.74\sigma$. We visually inspected the POSS I and II photographic plate scans to investigate the cause of these differences in proper motion. We find all but NLTT 43473 and NLTT 20123 to have incorrect J2000 position measurements and all but NLTT 43473 have spurious proper motion values upon comparison to other proper motion catalogues. Pearson product-moment correlation coefficients are 0.998, 0.995 and 0.988 $\mu_{\alpha} \cos \delta$, $\mu_{\delta}$ and $\mu_{total}$ respectively after removal of the 6 sources as discussed below.

\subsubsection*{NLTT 43473}
In the J1 image the high proper motion source is overlapping another source to the north with a separation of 1."3, likely causing the centroid on the object at J1 to be skewed north and further causing the observed larger proper motion in declination. Proper motion in right ascension agrees comfortably. Since this is a genuine match and the rNLTT proper motion is corroborated by the LSPM catalogue we have included it in the comparison.

\subsubsection*{NLTT 22010}
No high proper motion object is observed during blinking of 3' $\times$ 3' UKIDSS images, in agreement with our proper motion results. We included the 2MASS image in blinking and still no high proper motion object is evident. No source in our catalogue has a similar proper motion within 1' of the given position of NLTT 22010. We can see no source with stated rNLTT 22010 proper motion when blinking 12' $\times$ 12' POSS-I and POSS-II images (with a 42 year baseline the total expected movement is 7.8" which should be clearly visible). Also note that this source is not present in LSPM match even though its area should be covered. This high proper motion source is therefore questionable and it has been removed from our comparison.

\subsubsection*{NLTT 21214}
Inspection of 1' $\times$ 1' UKIDSS images centred on the rNLTT J2000 position of NLTT 21214 reveals the UKIDSS source as clearly extended and no proper motion, in agreement with our catalogue and consistent with the source being extragalactic. We located NLTT 21214 approximately 1.'25 to the north east of the \citet{salim03} given J2000 position. Furthermore the magnitude of the rNLTT proper motion for this source (-75 and -174 $mas~yr^{-1}$ in $\mu_{\alpha} \cos \delta$ and $\mu_{\delta}$ respectively) is not in agreement with the LSPM catalogue (-114$\pm$8 and -217$\pm$8 $mas~yr^{-1}$ in $\mu_{\alpha} \cos \delta$ and $\mu_{\delta}$) or our own (-123$\pm$9 and -228$\pm$8 $mas~yr^{-1}$ in $\mu_{\alpha} \cos \delta$ and $\mu_{\delta}$). The original NLTT proper motion is closer (-104 and -195 $mas~yr^{-1}$ in $\mu_{\alpha} \cos \delta$ and $\mu_{\delta}$). This source has been removed from the comparison due to a suspected incorrect rNLTT proper motion.

\subsubsection*{NLTT 20123}
High proper motion evident on blinking of UKIDSS and POSS images, direction of proper motion is in agreement with rNLTT and our catalogue. A rough centroid on the source at both epochs using the Region tool in \textsc{DS9} and WCS coordinates gives proper motions of 54 and -102 $mas~yr^{-1}$ in RA and Dec respectively, consistent with our catalogue values. USNO-B1.0 and LSPM proper motion values are also consistent with our catalogue. No source in our catalogue has a similar proper motion within 2' of the given position of NLTT 20123. We suspect the rNLTT proper motion of this source is incorrect and have removed it from our comparison.

\subsubsection*{NLTT 18649}
We blinked POSS-I (R band) and POSS-II (IR) images with an epoch baseline of 48 years and located NLTT 18649 1.'4 south south west of \citet{salim03} J2000 location. We located NLTT 18649 in our catalogue with a proper motion not in agreement with rNLTT but agreeing well with LSPM and USNO-B1.0 values. We suspect the rNLTT proper motion of this source is incorrect and have removed it from our comparison.

\subsubsection*{NLTT 18692}
We located NLTT 18692 1.'25 south west west of \citet{salim03} J2000 location. The rNLTT proper motion for 18692 is inconsistent with the USNO-B1.0 and LSPM catalogue values and has been removed from this comparison as a result. The LSPM and USNO-B1.0 proper motion values agree well with those of our catalogue.

\subsubsection*{NLTT 19021}
We located NLTT 19021 1.'4 south south east of \citet{salim03} J2000 location. The rNLTT proper motion for 19021 is inconsistent with the USNO-B1.0 and LSPM catalogue values and has been removed from this comparison as a result. Note that the LSPM and USNO-B1.0 proper motion values agree very well with those of our catalogue.

\subsection{Testing the Relative to Absolute Correction}\label{reltoabstest}

\begin{figure}
\begin{center}
\begin{tabular}{c}
\epsfig{file=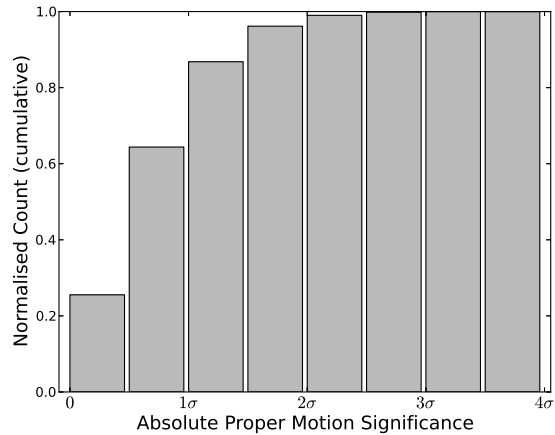,width=1\linewidth,clip=}
\end{tabular}
\caption{We identified 4,661 quasars within our catalogue using a method described in Section \ref{reltoabstest}. This plot shows the distribution of the proper motion significance of the quasar sample.}
\label{qsopmdist}
\end{center}
\end{figure}

We produced a list of quasar candidates by matching the full catalogue to the Large Quasar Reference Frame (LQRF; \citealt{andrei09}) using a 1" matching radius. We rejected quasar matches with more than one ULAS source within 3" and any that did not meet the restrictions imposed on reference stars described in Section \ref{cppselection}, leaving 4,661 quasar candidates. The mean absolute proper motion of this sample is $-0.44\pm{}0.16$ and $-0.08\pm{}0.15$ $mas~yr^{-1}$ in $\alpha \cos \delta$ and $\delta$ respectively. While the mean absolute proper motion of this sample in $\alpha \cos \delta$ is significant at the 2.75$\sigma$ level, we note that it is much smaller than the typical uncertainties on the proper motions (see Figure \ref{pmerrs}). Figure \ref{qsopmdist} shows that distribution of proper motion significances for this sample.

\begin{figure}
\begin{center}
\begin{tabular}{c}
\epsfig{file=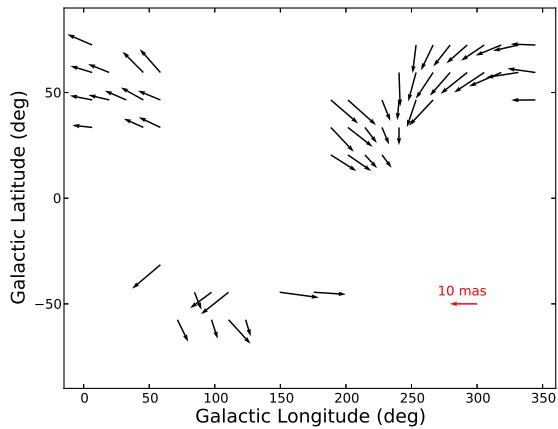,width=1\linewidth,clip=}
\end{tabular}
\caption{Median proper motions in galactic coordinates of a sample of 214,593 sources with well measured proper motions. The sample was separated into 13 $\times$ 13 degree ($l~\times~b$) bins. The red arrow shows an example motion of magnitude 10 mas. The points with Galactic latitude $b<0^{\circ}$ are those of isolated fields which suffer from a relative lack of reference galaxies.}
\label{r2aquiver}
\end{center}
\end{figure}

We selected a sample of 214,593 sources with which to test the direction and magnitude of the relative to absolute correction. Sources were selected in absolute proper motion space such that their motions were greater than three times their error and less than 500 $mas~yr^{-1}$, since we wanted to exclude the nearest sources, for which random velocity dispersion is the dominant factor in their proper motion, as opposed to Galactic location. We also selected only sources with 16 $<$ J1 \& J2 $<$ 19.6, J1 \& J2 uncertainty $<$ 0.1, J1 \& J2 ellipticity $<$ 0.3, and classified as stellar at both epochs.
We binned the sample in 13 $\times$ 13 degree bins ($l~\times~b$), rejecting any bins containing fewer than five sources. Proper motions were converted into the galactic coordinate system and the median motion of each bin was calculated. Figure \ref{r2aquiver} shows the median motions in galactic coordinates, which agrees well with an equivalent plot derived from Hipparcos measurements in \citeauthor{abad03} (\citeyear{abad03}; their figure 4). The points with Galactic latitude $b<0^{\circ}$ are those of the isolated fields which have very low relative to absolute correction reference galaxy counts (see Figure \ref{r2acnts}).
While the UKIDSS LAS and hence this catalogue were not designed to improve on the values of Oort's constants, using our sample we derive a value of $-13.79\pm{}6.58~km~s^{-1}~kpc^{-1}$ for the B constant. This agrees with a value of $-12.37\pm{}0.64~km~s^{-1}~kpc^{-1}$ derived from Hipparcos measurements by \citet{feast97} and should therefore validate our relative to absolute correction. The A constant in our case is related to radial velocity and hence a well constrained A constant is difficult to obtain.

\subsection{Investigation of Faint Limit}\label{reliability}

To attempt to quantify the reliability of catalogue proper motions and provide a reliable sample of brown dwarf candidates for binary searches (see Section \ref{blinkingbinaries}) we blinked all 980 sources with motions of 500 $mas$ or more that also met the following criteria:
\begin{description}
\item Y-J $>$ 0.7
\item J1 \& J2 ellipticity $<$ 0.3
\item J1 \& J2 classification $-1$ (stellar)
\item mergedClass $=$ $-1$ (stellar)
\end{description}
We assigned classifications of genuine, false and unsure based on their calculated motion compared to their apparent motion. A further classification, interesting was applied if there appeared to be another object in the 1'$\times$1' field with roughly the same motion vector (see Section \ref{blinkingbinaries}). We chose a minimum motion between epochs of 500 $mas$ since a motion of this magnitude should be detectable by eye, covering 2.5 pixels between the LAS J band images. 

First epoch images were obtained using the multiGetImage tool of the WSA and we wrote Linux scripts to retrieve second epoch J band images via the WSA Archive Listing tool. A further shell script was used to automatically select pairs of images and blink them using \textsc{DS9}, which made visual inspection of this sample of almost one thousand images possible in under a day. 

\begin{figure}
\begin{center}
\begin{tabular}{c}
\epsfig{file=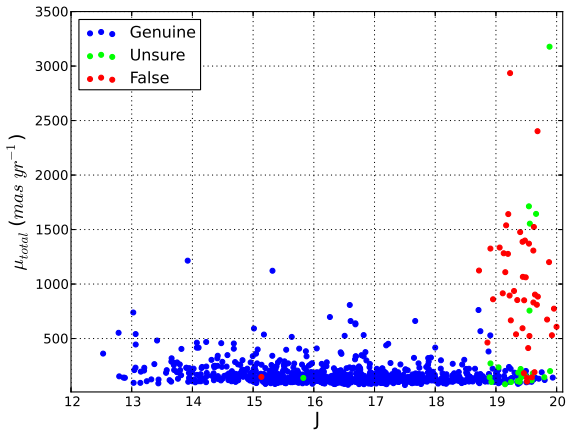,width=1\linewidth,clip=}
\end{tabular}
\caption{The distribution of {\it genuine}, {\it false} and {\it ambiguous} high proper motion candidates from Section \ref{reliability} in proper motion and J band brightness.}
\label{reliability_scatter}
\end{center}
\end{figure}

Figure \ref{reliability_scatter} shows the distribution of genuine, false and ambiguous proper motions for this sample in proper motion and brightness. We find the catalogue to be very reliable for red sources brighter than J$=$19. Reliability is diminished at the faint end but there are still many genuine high proper motion sources that can be found. We find a total of 834 genuine high proper motion sources in this sample. We note that the vast majority of false high proper motions were due to mismatched sources, which is to be expected due to the 0.5 J mag variability tolerance and given the increase in source density towards the faint end. 

\subsection{Catalogue Caveats}\label{catcaveats}

We find that ULAS sources brighter than 12 in J are often either saturated or very close to saturation and their centroids often fall in different places at different epochs and wavebands. This causes false high proper motions and large differences between the WSA proper motion values and ours. The vast majority of saturated objects were identified and flagged by the WSA and then removed from our catalogue by us, though further investigation showed that a few remained and as a result we elected to remove all sources brighter than 12 in the J band. 
We also find diffraction spikes of very bright stars as false high proper motion objects. Where these are not identified by the CASU/WSA pipelines they are usually identifiable as having large ellipticities and are easy to screen for through visual inspection, the pdf document report generated by the multiGetImage tool of the WSA in standard mode is sufficient in most cases.

Ideally quasars should be used for a relative to absolute correction but we would require several well distributed about each frame. A simple 1" match to the Large Quasar Reference Frame (LQRF; \citealt{andrei09}) yields one quasar for every two LAS frames on average which is insufficient for our purposes. Therefore we used galaxies as described in Section \ref{rel2abscorr}.

Relative proper motions are relative to the average motion of the reference stars used to compute the polynomial transform. Where a local transform is used the zero point motion is never exactly the same. This may introduce a small systematic random error into the absolute proper motion since the correction vectors are applied globally. Steps have been taken to limit this: only sources with small preliminarily measured residuals are used as reference sources in the final fit; and the requirement of at least 3 reference sources in each quadrant means that a minimum of of 12 reference sources are used. This should be sufficient to reduce any scatter in global - local mean motions. Indeed, we find that the mean difference between global and local residuals for bright stars is 13 $mas$ on each axis, which is only 20\% of the typical uncertainty on the residuals. Visual inspection of the spatial distribution of local-global variation shows no serious anomalies.

Sources LSPM J1625+2519 and LSPM J1609+2457 were found in our catalogue with proper motion measurements inconsistent with those of the LSPM catalogue (see Section \ref{sec:lspmcomp}). An inspection of the J band UKIDSS images indicated that the source of these inconsistent proper motions may be a poor quality UKIDSS frame for each leading to an inaccurate centroids on the sources at those epochs. This is probably due to the inclusion of a small number of poor quality UKIDSS images because much of the second epoch data have not yet been through the UKIDSS quality control procedures. We note however that 99.5\% of sources compared were unaffected by this and it is as such a minor issue.

\section{Results}\label{sec:results}

To capitalise on the availability of proper motions and a wide range of photometry for a large fraction of the LAS field we undertook several searches for new high proper motion objects which we detail here. Results of searches for new benchmark ultracool dwarfs can be found in Section \ref{newbenchmarkcands}. Note that unless stated otherwise Y, J, H, and K magnitudes in this section are on the MKO system and J band photometry is UKIDSS first epoch. We give first epoch J magnitude since it is most often contemporaneous with the Y band observation.

\subsection{Initial Searches for Interesting High Proper Motion Objects}\label{sec:hpm}

In a further effort to gauge the reliability of the catalogue to search for new high proper motion objects we selected a group of bright high proper motion objects from the 300 deg$^2$ of overlap with UKIDSS DR10 that also met the following restrictions:
\begin{description}
\item J1 $<$ 18
\item J1 \& J2 ellipticity $<$ 0.3
\item J1 \& J2 classification $-1$ (stellar)
\item total proper motion $>$ 300 $mas~yr^{-1}$
\end{description}
Note that there were no colour constraints in this selection. The selection left us with 42 sources to investigate. We retrieved their first and second epoch J band FITS images from the WSA using the multiGetImage tool and cross matched with SIMBAD to get names and alternative proper motions where available. We also cross matched to the SDSS ninth data release, which we verified visually, to retrieve ugriz optical photometry. Their images were blinked to verify their high proper motions. The values determined are given in Tables \ref{hpmastr} and \ref{hpmphot}. We note that one source is false (discussed below) and is therefore not included in these tables.

\begin{table*}
\begin{tabular}{|l|c|c|c|c|c|c|l|}
\hline
  \multicolumn{1}{|c|}{Name} &
  \multicolumn{1}{c|}{$\alpha_{J2000}$} &
  \multicolumn{1}{c|}{$\delta_{J2000}$} &
  \multicolumn{1}{c|}{$\mu_{\alpha} \cos \delta$} &
  \multicolumn{1}{c|}{$\mu_{\delta}$} &
  \multicolumn{1}{c|}{alt. $\mu_{\alpha} \cos \delta$} &
  \multicolumn{1}{c|}{alt. $\mu_{\delta}$} &
  \multicolumn{1}{c|}{alt. $\mu$ source} \\
\hline
  LP 365-11 & 07:28:25.75 & +24:31:51.9 & $136 \pm 6$ & $-315 \pm 6$ & $148 \pm 8$ & $-321 \pm 8$ & LSPM\\
  LP 65-25 & 07:35:02.85 & +24:57:04.4 & $201 \pm 5$ & $-251 \pm 6$ & $199 \pm 8$ & $-238 \pm 8$ & LSPM\\
  2MASS J07414920+2351275 & 07:41:49.18 & +23:51:27.8 & $-262 \pm 11$ & $-212 \pm 9$ & $-250 \pm 12$ & $-116 \pm 13$ & $^a$\\
  LSR J0745+2627 & 07:45:08.95 & +26:27:06.4 & $527 \pm 12$ & $-719 \pm 13$ & $496 \pm 8$ & $-744 \pm 8$ & LSPM\\
  2MASS J07474639+2605167 & 07:47:46.39 & +26:05:17.5 & $-189 \pm 9$ & $-245 \pm 10$ & $-253 \pm 49$ & $-170 \pm 50$ & $^b$\\
  LP 366-18 & 07:49:17.18 & +21:03:35.8 & $69 \pm 9$ & $-304 \pm 9$ & $65 \pm 8$ & $-299 \pm 8$ & LSPM\\
  LHS 1953 & 07:52:08.12 & +27:00:01.5 & $609 \pm 7$ & $-658 \pm 8$ & $604 \pm 8$ & $-667 \pm 8$ & LSPM\\
  LP 366-27 & 07:56:40.85 & +23:36:35.6 & $74 \pm 7$ & $-305 \pm 7$ & $93 \pm 8$ & $-307 \pm 8$ & LSPM\\
  2MASS J08044064+2239502 & 08:04:40.63 & +22:39:49.7 & $12 \pm 10$ & $-320 \pm 11$ & $4 \pm 8$ & $-336 \pm 8$ & LSPM\\
  LP 424-14 & 08:09:40.24 & +19:32:04.3 & $-396 \pm 7$ & $-109 \pm 6$ & $-398 \pm 8$ & $-110 \pm 8$ & LSPM\\
  ULAS J081045.24+222841.9 & 08:10:45.25 & +22:28:44.1 & $-20 \pm 10$ & $-306 \pm 8$ & ... & ... & ...\\
  ULAS J081127.84+203925.7 & 08:11:27.82 & +20:39:28.4 & $40 \pm 8$ & $-460 \pm 8$ & ... & ... & ...\\
  LHS 6139 & 08:11:27.90 & +20:39:26.2 & $32 \pm 8$ & $-461 \pm 8$ & $37 \pm 8$ & $-467 \pm 8$ & LSPM\\
  G 40-12 & 08:13:24.20 & +26:57:10.6 & $351 \pm 11$ & $-253 \pm 7$ & $340 \pm 8$ & $-259 \pm 8$ & LSPM\\
  LP 367-56 & 08:16:36.29 & +23:06:16.1 & $96 \pm 8$ & $-344 \pm 5$ & $84 \pm 8$ & $-350 \pm 8$ & LSPM\\
  EGGR 531 & 08:16:42.05 & +21:37:36.0 & $-93 \pm 9$ & $-397 \pm 6$ & $-104 \pm 8$ & $-392 \pm 8$ & LSPM\\
  ULAS J082155.56+250939.8 & 08:21:55.79 & +25:09:40.2 & $-448 \pm 11$ & $-62 \pm 14$ & ... & ... & ...\\
  LHS 2006 & 08:23:47.97 & +24:56:57.7 & $237 \pm 6$ & $-479 \pm 7$ & $235 \pm 8$ & $-471 \pm 8$ & LSPM\\
  2MASS J08253258+2359306 & 08:25:32.59 & +23:59:30.6 & $-6 \pm 7$ & $-327 \pm 7$ & $15 \pm 8$ & $-320 \pm 8$ & LSPM\\
  LP 311-21 & 08:28:35.05 & +26:45:33.1 & $193 \pm 11$ & $-251 \pm 11$ & $199 \pm 8$ & $-239 \pm 8$ & LSPM\\
  2MASS J08332144+2300120 & 08:33:21.45 & +23:00:11.8 & $65 \pm 7$ & $-314 \pm 10$ & $72 \pm 8$ & $-319 \pm 8$ & LSPM\\
  LSPM J0836+2432 & 08:36:18.07 & +24:32:56.7 & $238 \pm 10$ & $-499 \pm 13$ & $231 \pm 8$ & $-496 \pm 8$ & LSPM\\
  LP 321-30 & 08:46:01.27 & +27:23:07.5 & $-108 \pm 11$ & $-447 \pm 5$ & $-103 \pm 8$ & $-443 \pm 8$ & LSPM\\
  ULAS J085335.33+285902.4 & 08:53:35.59 & +28:59:07.0 & $-471 \pm 20$ & $-629 \pm 11$ & ... & ... & ...\\
  LP 260-3 & 09:16:06.52 & +32:56:03.0 & $-229 \pm 8$ & $-238 \pm 7$ & $-236 \pm 8$ & $-229 \pm 8$ & LSPM\\
  LP 313-36 & 09:17:43.21 & +30:56:50.9 & $-23 \pm 8$ & $-304 \pm 6$ & $-21 \pm 8$ & $-306 \pm 8$ & LSPM\\
  WD 0921+315 & 09:24:30.86 & +31:20:33.6 & $-204 \pm 13$ & $-369 \pm 10$ & $-193 \pm 8$ & $-378 \pm 8$ & LSPM\\
  2MASS J15052821+3115037 & 15:05:28.21 & +31:15:02.9 & $-20 \pm 6$ & $-512 \pm 7$ & $-37$ & $-529$ & $^c$\\
  LP 272-48 & 15:10:38.43 & +33:10:16.9 & $-43 \pm 7$ & $-361 \pm 7$ & $-45 \pm 8$ & $-365 \pm 8$ & LSPM\\
  LP 327-24 & 15:11:51.21 & +30:33:06.2 & $-397 \pm 8$ & $-283 \pm 8$ & $-393 \pm 8$ & $-265 \pm 8$ & LSPM\\
  ULAS J151354.98+303543.9 & 15:13:54.91 & +30:35:46.2 & $156 \pm 8$ & $-421 \pm 9$ & ... & ... & ...\\
  LHS 3042 & 15:14:26.02 & +30:23:34.0 & $-583 \pm 9$ & $-9 \pm 7$ & $-603 \pm 8$ & $-5 \pm 8$ & LSPM\\
  LHS 3063 & 15:21:51.72 & +30:48:26.2 & $-412 \pm 8$ & $341 \pm 8$ & $-413 \pm 8$ & $339 \pm 8$ & LSPM\\
  2MASS J15593876+2550362 & 15:59:38.80 & +25:50:36.3 & $-358 \pm 10$ & $108 \pm 10$ & $-328 \pm 37$ & $119 \pm 37$ & $^b$\\
  ULAS J160036.59+284305.7 & 16:00:36.70 & +28:43:04.2 & $-228 \pm 13$ & $228 \pm 12$ & ... & ... & ...\\
  NLTT 41963 & 16:05:52.82 & +25:11:38.8 & $-337 \pm 7$ & $-5 \pm 7$ & $-339 \pm 8$ & $-4 \pm 8$ & LSPM\\
  NLTT 42004 & 16:06:35.73 & +24:28:40.9 & $-93 \pm 5$ & $-309 \pm 4$ & $-86 \pm 8$ & $-325 \pm 8$ & LSPM\\
  NLTT 42650 & 16:22:40.15 & +29:19:13.0 & $-298 \pm 8$ & $-226 \pm 9$ & $-296 \pm 8$ & $-218 \pm 8$ & LSPM\\
  LP 330-15 & 16:26:24.56 & +28:56:26.0 & $-151 \pm 9$ & $-304 \pm 9$ & $-152 \pm 8$ & $-298 \pm 8$ & LSPM\\
  LHS 3198 & 16:27:40.18 & +29:27:15.1 & $-173 \pm 7$ & $-532 \pm 8$ & $-156 \pm 8$ & $-537 \pm 8$ & LSPM\\
  LSPM J1641+3210 & 16:41:43.41 & +32:10:39.0 & $-350 \pm 6$ & $29 \pm 6$ & $-370 \pm 8$ & $37 \pm 8$ & LSPM\\
\hline
\multicolumn{8}{|l|}{$^a$ \citet{casewell08}}\\
\multicolumn{8}{|l|}{$^b$ \citet{zhang10}}\\
\multicolumn{8}{|l|}{$^c$ \citet{sheppard09}, stated total proper motion uncertainty is about 10\%}\\
\end{tabular}
\caption{\textbf{Astrometry} Astrometry for 41 genuine high proper motion ($> 300 mas~yr^{-1}$) sources from our proper motion catalogue. Coordinates are those of the UKIDSS LAS (J1 epoch), converted to 2000.0 epoch using the stated proper motion values. Proper motions are given in units of $mas~yr^{-1}$.}
\label{hpmastr}
\end{table*}

\begin{landscape}
\begin{table}
\centering
\begin{tabular}{|l|c|c|c|c|c|c|c|c|c|c|}
\hline
  \multicolumn{1}{|c|}{Name} &
  \multicolumn{1}{c|}{u} &
  \multicolumn{1}{c|}{g} &
  \multicolumn{1}{c|}{r} &
  \multicolumn{1}{c|}{i} &
  \multicolumn{1}{c|}{z} &
  \multicolumn{1}{c|}{Y} &
  \multicolumn{1}{c|}{J} &
  \multicolumn{1}{c|}{H} &
  \multicolumn{1}{c|}{K} \\
\hline
  LP 365-11 & $ 20.59 \pm 0.05 $ & $ 17.98 \pm 0.01 $ & $ 16.59 \pm 0.01 $ & $ 16.02 \pm 0.01 $ & $ 15.73 \pm 0.01 $ & $ 14.980 \pm 0.004 $ & $ 14.525 \pm 0.010 $ & $ 14.043 \pm 0.004 $ & $ 13.865 \pm 0.004 $\\
  LP 65-25 & $ 20.17 \pm 0.05 $ & $ 17.62 \pm 0.01 $ & $ 16.24 \pm 0.00 $ & $ 15.67 \pm 0.00 $ & $ 15.34 \pm 0.01 $ & $ 14.581 \pm 0.003 $ & $ 14.178 \pm 0.010 $ & $ 13.673 \pm 0.003 $ & $ 13.448 \pm 0.004 $\\
  2MASS J07414920+2351275 & $ 25.44 \pm 0.63 $ & $ 24.31 \pm 0.43 $ & $ 24.34 \pm 0.52 $ & $ 24.85 \pm 0.41 $ & $ 19.56 \pm 0.06 $ & $ 17.129 \pm 0.013 $ & $ 15.880 \pm 0.010 $ & $ 16.104 \pm 0.023 $ & $ 16.277 \pm 0.037 $\\
  LSR J0745+2627 & $ 22.65 \pm 0.24 $ & $ 19.99 \pm 0.02 $ & $ 18.69 \pm 0.01 $ & $ 18.23 \pm 0.01 $ & $ 17.96 \pm 0.02 $ & $ 17.389 \pm 0.014 $ & $ 17.122 \pm 0.013 $ & $ 17.087 \pm 0.051 $ & $ 17.184 \pm 0.080 $\\
  2MASS J07474639+2605167 & $ 25.35 \pm 0.68 $ & $ 25.37 \pm 0.56 $ & $ 23.86 \pm 0.42 $ & $ 20.68 \pm 0.04 $ & $ 18.86 \pm 0.03 $ & $ 17.547 \pm 0.016 $ & $ 16.679 \pm 0.011 $ & $ 16.182 \pm 0.020 $ & $ 15.778 \pm 0.025 $\\
  LP 366-18 & $ 21.78 \pm 0.14 $ & $ 19.49 \pm 0.01 $ & $ 18.07 \pm 0.01 $ & $ 17.51 \pm 0.01 $ & $ 17.17 \pm 0.01 $ & $ 16.448 \pm 0.008 $ & $ 15.965 \pm 0.010 $ & $ 15.531 \pm 0.009 $ & $ 15.375 \pm 0.012 $\\
  LHS 1953 & $ 20.29 \pm 0.04 $ & $ 17.50 \pm 0.01 $ & $ 15.93 \pm 0.00 $ & $ 15.27 \pm 0.00 $ & $ 14.88 \pm 0.01 $ & $ 14.106 \pm 0.002 $ & $ 13.648 \pm 0.010 $ & $ 13.228 \pm 0.002 $ & $ 13.027 \pm 0.003 $\\
  LP 366-27 & $ 21.79 \pm 0.16 $ & $ 19.24 \pm 0.01 $ & $ 17.79 \pm 0.01 $ & $ 16.46 \pm 0.00 $ & $ 15.75 \pm 0.01 $ & $ 14.896 \pm 0.004 $ & $ 14.376 \pm 0.010 $ & $ 13.934 \pm 0.004 $ & $ 13.652 \pm 0.005 $\\
  2MASS J08044064+2239502 & $ 19.78 \pm 0.03 $ & $ 18.28 \pm 0.01 $ & $ 17.60 \pm 0.01 $ & $ 17.39 \pm 0.01 $ & $ 17.31 \pm 0.01 $ & $ 16.818 \pm 0.011 $ & $ 16.693 \pm 0.010 $ & $ 16.832 \pm 0.034 $ & $ 17.238 \pm 0.072 $\\
  LP 424-14 & $ 20.72 \pm 0.06 $ & $ 18.07 \pm 0.01 $ & $ 16.62 \pm 0.00 $ & $ 15.08 \pm 0.00 $ & $ 14.26 \pm 0.01 $ & $ 13.332 \pm 0.002 $ & $ 12.761 \pm 0.010 $ & $ 12.365 \pm 0.001 $ & $ 11.997 \pm 0.002 $\\
  ULAS J081045.24+222841.9 & $ 25.41 \pm 0.72 $ & $ 22.20 \pm 0.09 $ & $ 20.28 \pm 0.03 $ & $ 18.82 \pm 0.01 $ & $ 18.02 \pm 0.02 $ & $ 17.189 \pm 0.013 $ & $ 16.605 \pm 0.010 $ & $ 16.185 \pm 0.016 $ & $ 15.952 \pm 0.023 $\\
  ULAS J081127.84+203925.7 & $ 22.67 \pm 0.56 $ & $ 18.48 \pm 0.02 $ & $ 17.49 \pm 0.01 $ & $ 17.76 \pm 0.02 $ & $ 16.98 \pm 0.02 $ & $ 15.999 \pm 0.007 $ & $ 15.519 \pm 0.010 $ & $ 15.110 \pm 0.008 $ & $ 14.908 \pm 0.012 $\\
  LHS 6139 & $ 17.77 \pm 0.01 $ & $ 15.84 \pm 0.00 $ & $ 15.73 \pm 0.01 $ & $ 14.25 \pm 0.00 $ & $ 13.98 \pm 0.00 $ & $ 12.874 \pm 0.001 $ & $ 12.448 \pm 0.010 $ & $ 11.996 \pm 0.001 $ & $ 11.778 \pm 0.001 $\\
  G 40-12 & $ 18.15 \pm 0.01 $ & $ 15.82 \pm 0.00 $ & $ 14.69 \pm 0.00 $ & $ 14.83 \pm 0.01 $ & $ 13.97 \pm 0.00 $ & $ 13.218 \pm 0.002 $ & $ 12.812 \pm 0.010 $ & $ 12.292 \pm 0.001 $ & $ 12.111 \pm 0.002 $\\
  LP 367-56 & $ 21.56 \pm 0.12 $ & $ 18.81 \pm 0.01 $ & $ 17.33 \pm 0.01 $ & $ 16.76 \pm 0.01 $ & $ 16.40 \pm 0.01 $ & $ 15.674 \pm 0.005 $ & $ 15.217 \pm 0.010 $ & $ 14.776 \pm 0.006 $ & $ 14.613 \pm 0.008 $\\
  EGGR 531 & $ 17.95 \pm 0.01 $ & $ 17.18 \pm 0.00 $ & $ 16.84 \pm 0.00 $ & $ 16.71 \pm 0.00 $ & $ 16.70 \pm 0.01 $ & $ 16.168 \pm 0.006 $ & $ 15.954 \pm 0.010 $ & $ 15.728 \pm 0.009 $ & $ 15.665 \pm 0.016 $\\
  ULAS J082155.56+250939.8 & ... & ... & ... & ... & ... & $ 18.610 \pm 0.043 $ & $ 17.226 \pm 0.015 $ & $ 17.290 \pm 0.065 $ & $ 17.232 \pm 0.095 $\\
  LHS 2006 & $ 21.78 \pm 0.14 $ & $ 18.72 \pm 0.01 $ & $ 17.08 \pm 0.00 $ & $ 16.08 \pm 0.00 $ & $ 15.54 \pm 0.01 $ & $ 14.739 \pm 0.003 $ & $ 14.254 \pm 0.010 $ & $ 13.791 \pm 0.003 $ & $ 13.534 \pm 0.004 $\\
  2MASS J08253258+2359306 & $ 22.68 \pm 0.26 $ & $ 19.79 \pm 0.02 $ & $ 18.31 \pm 0.01 $ & $ 16.53 \pm 0.00 $ & $ 15.59 \pm 0.01 $ & $ 14.604 \pm 0.003 $ & $ 14.068 \pm 0.010 $ & $ 14.14 \pm 0.03 $$~^a$ & $ 13.336 \pm 0.004 $\\
  LP 311-21 & $ 21.30 \pm 0.10 $ & $ 18.62 \pm 0.01 $ & $ 17.15 \pm 0.00 $ & $ 16.52 \pm 0.00 $ & $ 16.13 \pm 0.01 $ & $ 15.359 \pm 0.005 $ & $ 14.934 \pm 0.010 $ & $ 14.430 \pm 0.006 $ & $ 14.210 \pm 0.008 $\\
  2MASS J08332144+2300120 & $ 21.95 \pm 0.16 $ & $ 19.03 \pm 0.01 $ & $ 17.52 \pm 0.01 $ & $ 15.80 \pm 0.00 $ & $ 14.88 \pm 0.01 $ & $ 13.903 \pm 0.002 $ & $ 13.359 \pm 0.010 $ & $ 12.923 \pm 0.002 $ & $ 12.614 \pm 0.002 $\\
  LSPM J0836+2432 & $ 20.30 \pm 0.05 $ & $ 19.51 \pm 0.01 $ & $ 18.90 \pm 0.01 $ & $ 18.74 \pm 0.01 $ & $ 18.69 \pm 0.03 $ & $ 18.197 \pm 0.026 $ & $ 17.985 \pm 0.028 $ & $ 17.909 \pm 0.096 $ & $ 17.968 \pm 0.162 $\\
  LP 321-30 & $ 21.93 \pm 0.16 $ & $ 19.01 \pm 0.01 $ & $ 17.33 \pm 0.00 $ & $ 16.44 \pm 0.00 $ & $ 15.95 \pm 0.01 $ & $ 15.136 \pm 0.004 $ & $ 14.643 \pm 0.010 $ & $ 14.176 \pm 0.004 $ & $ 13.906 \pm 0.006 $\\
  ULAS J085335.33+285902.4 & $ 23.16 \pm 0.50 $ & $ 24.12 \pm 0.43 $ & $ 24.09 \pm 0.62 $ & $ 22.17 \pm 0.20 $ & $ 20.33 \pm 0.13 $ & $18.930 \pm 0.129 $ & $ 17.705 \pm 0.031 $ & $ 16.936 \pm 0.123 $ & $ 16.425 \pm 0.044 $\\
  LP 260-3 & $ 22.84 \pm 0.28 $ & $ 19.81 \pm 0.02 $ & $ 18.20 \pm 0.01 $ & $ 17.26 \pm 0.01 $ & $ 16.69 \pm 0.01 $ & $ 15.891 \pm 0.006 $ & $ 15.403 \pm 0.010 $ & $ 14.931 \pm 0.010 $ & $ 14.670 \pm 0.012 $\\
  LP 313-36 & $ 21.86 \pm 0.13 $ & $ 19.20 \pm 0.01 $ & $ 17.61 \pm 0.01 $ & $ 16.80 \pm 0.01 $ & $ 16.33 \pm 0.01 $ & $ 15.537 \pm 0.004 $ & $ 15.041 \pm 0.010 $ & $ 14.591 \pm 0.005 $ & $ 14.323 \pm 0.007 $\\
  WD 0921+315 & $ 20.66 \pm 0.06 $ & $ 18.73 \pm 0.01 $ & $ 17.93 \pm 0.01 $ & $ 17.64 \pm 0.01 $ & $ 17.50 \pm 0.02 $ & $ 16.927 \pm 0.009 $ & $ 16.631 \pm 0.010 $ & $ 16.408 \pm 0.022 $ & $ 16.401 \pm 0.038 $\\
  2MASS J15052821+3115037 & $ 23.27 \pm 0.42 $ & $ 20.77 \pm 0.03 $ & $ 19.09 \pm 0.01 $ & $ 17.71 \pm 0.01 $ & $ 16.96 \pm 0.01 $ & $ 16.131 \pm 0.007 $ & $ 15.547 \pm 0.010 $ & $ 15.115 \pm 0.007 $ & $ 14.810 \pm 0.010 $\\
  LP 272-48 & $ 20.91 \pm 0.06 $ & $ 18.15 \pm 0.01 $ & $ 16.67 \pm 0.01 $ & $ 15.35 \pm 0.01 $ & $ 14.62 \pm 0.01 $ & $ 13.768 \pm 0.002 $ & $ 13.239 \pm 0.010 $ & $ 12.811 \pm 0.002 $ & $ 12.555 \pm 0.002 $\\
  LP 327-24 & $ 21.62 \pm 0.11 $ & $ 19.18 \pm 0.01 $ & $ 17.81 \pm 0.01 $ & $ 15.81 \pm 0.00 $ & $ 14.63 \pm 0.00 $ & $ 13.479 \pm 0.002 $ & $ 12.843 \pm 0.010 $ & $ 12.259 \pm 0.001 $ & $ 11.854 \pm 0.001 $\\
  ULAS J151354.98+303543.9 & $ 24.60 \pm 0.76 $ & $ 21.44 \pm 0.04 $ & $ 20.05 \pm 0.02 $ & $ 17.31 \pm 0.01 $ & $ 15.79 \pm 0.01 $ & $ 14.543 \pm 0.003 $ & $ 13.850 \pm 0.010 $ & $ 13.343 \pm 0.002 $ & $ 12.911 \pm 0.002 $\\
  LHS 3042 & $ 23.30 \pm 0.42 $ & $ 20.36 \pm 0.02 $ & $ 18.72 \pm 0.01 $ & $ 16.90 \pm 0.01 $ & $ 15.92 \pm 0.01 $ & $ 14.973 \pm 0.004 $ & $ 14.401 \pm 0.010 $ & $ 14.015 \pm 0.003 $ & $ 13.697 \pm 0.004 $\\
  LHS 3063 & $ 20.25 \pm 0.05 $ & $ 18.01 \pm 0.01 $ & $ 16.52 \pm 0.00 $ & $ 15.03 \pm 0.00 $ & $ 14.21 \pm 0.00 $ & $ 13.260 \pm 0.001 $ & $ 12.763 \pm 0.010 $ & $ 12.344 \pm 0.001 $ & $ 12.043 \pm 0.001 $\\
  2MASS J15593876+2550362 & $ 24.86 \pm 0.59 $ & $ 24.38 \pm 0.33 $ & $ 23.71 \pm 0.30 $ & $ 20.43 \pm 0.03 $ & $ 18.57 \pm 0.03 $ & $ 17.293 \pm 0.017 $ & $ 16.442 \pm 0.010 $ & $ 16.099 \pm 0.014 $ & $ 15.647 \pm 0.018 $\\
  ULAS J160036.59+284305.7 & $ 23.95 \pm 0.79 $ & $ 23.82 \pm 0.35 $ & $ 23.51 \pm 0.38 $ & $ 21.89 \pm 0.16 $ & $ 21.49 \pm 0.42 $ & $ 18.856 \pm 0.060 $ & $ 17.650 \pm 0.029 $ & $ 16.833 \pm 0.025 $ & $ 16.136 \pm 0.031 $\\
  NLTT 41963 & $ 20.42 \pm 0.05 $ & $ 18.12 \pm 0.01 $ & $ 16.66 \pm 0.01 $ & $ 15.17 \pm 0.00 $ & $ 14.37 \pm 0.01 $ & $ 13.393 \pm 0.002 $ & $ 12.897 \pm 0.010 $ & $ 12.481 \pm 0.001 $ & $ 12.179 \pm 0.002 $\\
  NLTT 42004 & $ 22.35 \pm 0.17 $ & $ 19.46 \pm 0.01 $ & $ 17.83 \pm 0.01 $ & $ 16.96 \pm 0.01 $ & $ 16.45 \pm 0.01 $ & $ 15.596 \pm 0.005 $ & $ 15.131 \pm 0.010 $ & $ 14.638 \pm 0.004 $ & $ 14.406 \pm 0.008 $\\
  NLTT 42650 & $ 21.99 \pm 0.16 $ & $ 19.85 \pm 0.01 $ & $ 18.88 \pm 0.01 $ & $ 18.49 \pm 0.01 $ & $ 18.33 \pm 0.02 $ & $ 17.743 \pm 0.022 $ & $ 17.462 \pm 0.022 $ & $ 17.355 \pm 0.039 $ & $ 17.305 \pm 0.083 $\\
  LP 330-15 & $ 24.10 \pm 0.76 $ & $ 20.04 \pm 0.02 $ & $ 18.39 \pm 0.01 $ & $ 17.11 \pm 0.01 $ & $ 16.47 \pm 0.01 $ & $ 15.579 \pm 0.005 $ & $ 15.050 \pm 0.010 $ & $ 14.612 \pm 0.005 $ & $ 14.312 \pm 0.006 $\\
  LHS 3198 & $ 19.73 \pm 0.04 $ & $ 17.10 \pm 0.00 $ & $ 15.62 \pm 0.00 $ & $ 15.06 \pm 0.00 $ & $ 14.71 \pm 0.00 $ & $ 13.923 \pm 0.002 $ & $ 13.498 \pm 0.010 $ & $ 13.016 \pm 0.002 $ & $ 12.837 \pm 0.002 $\\
  LSPM J1641+3210 & $ 23.22 \pm 0.38 $ & $ 20.74 \pm 0.03 $ & $ 19.22 \pm 0.01 $ & $ 16.95 \pm 0.01 $ & $ 15.71 \pm 0.01 $ & $ 14.642 \pm 0.003 $ & $ 14.012 \pm 0.010 $ & $ 13.574 \pm 0.003 $ & $ 13.223 \pm 0.003 $\\
\hline
\multicolumn{10}{|l|}{$^a$ No LAS H band data available, 2MASS H magnitude given.}\\
\end{tabular}
\caption{\textbf{Photometry} SDSS optical and UKIDSS near infrared photometry for 41 high proper motion ($>$ 300$~mas~yr^{-1}$) sources from our proper motion catalogue. J band magnitude is first epoch UKIDSS LAS.}
\label{hpmphot}
\end{table}
\end{landscape}

Here we itemise sources of interest amongst the 41 bright, high proper motion sources.

\subsubsection*{LSR J0745+2627}
This object was selected as one of the highest proper motion sources in a prototype version of this catalogue based on UKIDSS LAS DR9 data. It has previous proper motion measurements by \citet{lepine05} and white dwarf identification by \citet{reid03}. Using this catalogue LSR J0745+2627 was re-identified by \citet{catalan12} as the brightest pure-H ultracool ($T_{eff} < 4000~K$) white dwarf currently known.

\subsubsection*{LHS 6139 and ULAS J081127.84+203925.7}
\begin{figure}
\begin{center}
\begin{tabular}{c}
\epsfig{file=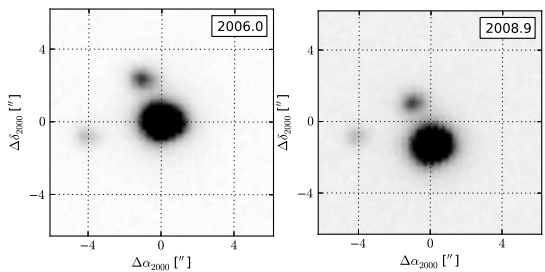,width=1\linewidth,clip=}
\end{tabular}
\caption{First and second epoch LAS J band images of LHS 6139 and ULAS J081127.84+203925.7 centred on the first epoch position of the former. Their common proper motion is evident.}
\label{lhs6139cpm}
\end{center}
\end{figure}

These objects share a common proper motion (see Figure \ref{lhs6139cpm}). The difference in their measured proper motions is half its uncertainty.

\subsubsection*{ULAS J082155.56+250939.8}
The T4.5 dwarf ULAS J082155.56+250939.8, confirmed with a NIRI spectrum, identified by \citet{burningham13}.

\subsubsection*{2MASS J07414920+2351275}
The T5 dwarf \citep{burgasser06} 2MASS J07414920+2351275 has a proper motion discrepancy between our catalogue and previous measurements by \citet{casewell08} ($-250.22\pm12.18~mas~yr^{-1}$ and $-116.21\pm13.32~mas~yr^{-1}$ in RA and Dec respectively) and \citet{faherty09} ($-243\pm13~mas~yr^{-1}$ and $-143\pm14~mas~yr^{-1}$ in RA and Dec respectively).  \citet{faherty09} also provide a distance of $18\pm2~pc$ for this object. There is no obvious defect present in the two J band images which might cause such an error in the proper motion. The relatively large parallax of the source can be ruled out as the source of proper motion error in our catalogue since our epoch baseline is 11 days from a year. We also find that the WSA proper motion, derived from detections in all five bands, is consistent with our value. The source of this discrepancy remains unknown.

\subsubsection*{2MASS J08044064+2239502 \& NLTT 42650}
Identified by \citet{kilic10}, the DZ White Dwarf 2MASS J08044064+2239502 and the DC White Dwarf NLTT 42650. 

\subsubsection*{EGGR 531}
The well studied DA8 White Dwarf EGGR 531, first identified by \citet{greenstein80}.

\subsubsection*{LP 260-3, 2MASS J15593876+2550362 \& LSPM J1641+3210}
The M2, M6 and M7 type dwarfs LP 260-3, LSPM J1641+3210 and 2MASS J15593876+2550362 are previously studied separate systems. Spectral types, photometric distances ($508 pc$, $55.9 pc$ and $161.9 pc$) and radial velocities ($105 km~s^{-1}$, $-6.1  km~s^{-1}$ and $-54.7 km~s^{-1}$) were measured by \citet{west08} using their respective SDSS DR5 spectra.\\

\subsubsection*{WD 0921+315}
The $4810\pm60K$ DC Helium rich White Dwarf WD 0921+315 identified by \citet{sayres12}. The SDSS spectrum provides spectroscopic confirmation.

\subsubsection*{ups Gem Ghost}
\begin{figure}
\begin{center}
\begin{tabular}{c}
\epsfig{file=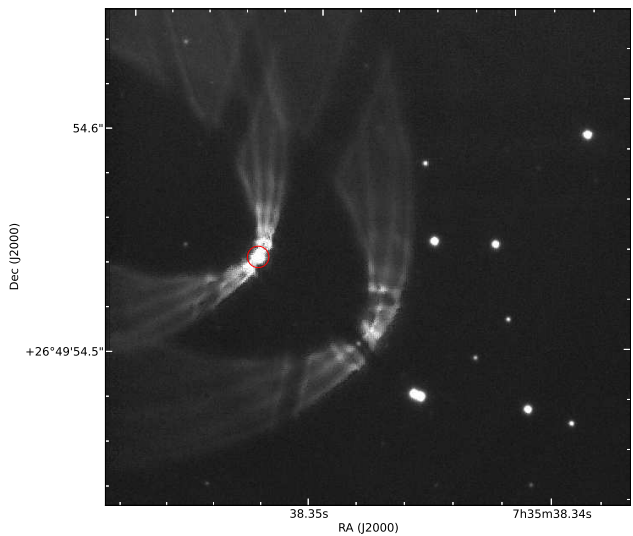,width=1\linewidth,clip=}
\end{tabular}
\caption{The first epoch J band image of the ups Gem ghost source, which has a different position at each epoch giving a false high proper motion. The source is indicated by the red circle.}
\label{upsgemghost}
\end{center}
\end{figure}

A ghosted image of ups Gem (see Figure \ref{upsgemghost}) is the only false high proper motion ($-224 \pm 10$ \& $-1547 \pm 9$ $mas~yr^{-1}$ in $\mu_{\alpha} \cos \delta$ and $\mu_{\delta}$ respectively) source to have escaped rejection by the cuts described above. Suggesting that while they are effective at removing a lot of false high proper motion sources some will remain. If a clean high proper motion sample is required then blinking the first and second epoch J band images is recommended where practical. Images may be retrieved and blinked quickly using the WSA MultiGetImage tool and an image viewer which accepts command line input such as \textsc{DS9}. 

\subsubsection*{ULAS J075015.48+203650.0}
This object was missing from the above selection due to its second J band epoch classification as a galaxy, but was identified in other searches. Based on the \citet{hawley02} $i-J$ colour to spectral type table and the source's SDSS DR7 i band magnitude of 21.21$\pm$0.09 (note that the source is missing from SDSS data releases 8 and 9), it is a candidate M6/M7 dwarf at a distance of between 260 and 370 $pc$. It has a $504\pm18$$~mas~yr^{-1}$ proper motion. This corresponds to a range of tangential velocities between 620 and 870 $km~s^{-1}$, above the Galactic escape velocity.

\subsection{White Dwarfs}\label{whitedwarfs}

Ultra cool white dwarfs are among the oldest objects in the galaxy. Their ages are often very well constrained due to their predictable cooling rate based on theoretical models (eg. \citealt{meng10}, \citealt{chen10}), dwarf mass to progenitor star mass relationship and main sequence progenitor lifetime. Hence, these objects are ideal tools for placing lower limits on the age of the Galaxy and can give us clues to the conditions of a young Milky Way \citep{kilic06a}. A number of cool white dwarfs have been discovered to date, the usual method of discovery is photometric and reduced proper motion selection (\citealt{kilic05}, \citealt{leggett11}), often followed by spectroscopic confirmation.

For identification of white dwarfs in this catalogue, a cross match with optical catalogues will be necessary. We identify white dwarf candidates using a combination of cuts on near infrared and optical (SDSS) colours, proper motion and reduced proper motion, and other selections based on classification and ellipticity designed to reject possible false positives. Our candidates are likely cool white dwarfs based on fits of their photometry to model spectra. We refer the reader to \citet{catalan12} for a description of the ultra-cool H rich white dwarf LSR J0745+2627 which was identified due to its unusually high proper motion in an early version of this catalogue and subsequent photometric analysis and spectroscopic confirmation.

\subsection{L dwarfs}\label{ldwarfs}

Several hundred L dwarfs have been identified in the local field by wide field surveys, e.g. \citeauthor{kirkpatrick08}(\citeyear{kirkpatrick99}; \citeyear{kirkpatrick08}). Detection of new L dwarfs remains valuable because only a few have been identified that are in age-benchmark binaries (e.g. \citealt{zhang10}) which can be used to test model atmospheres. Similarly, a very small proportion of L dwarfs  have high space motions and unusually red or unusually blue (J-K) colours (see e.g. \citealp{kirkpatrick10} and references therein). Some of these are sub-L dwarfs with low metallicity and halo kinematics ($v_{tan}=$200-320 $km~s^{-1}$, \citealt{schilbach09}). More detections of such rare types of L dwarf are needed to better understand the population as a whole, and it is reasonable to expect that a large proper motion survey such as this may find some examples.

We present here the initial results of a search of the UKIDSS DR8 subset of our proper motion catalogue in order to illustrate the practicality of future searches. The DR8 subset comprises 260 deg$^2$ of our catalogue.
We first selected sources whose proper motions have $\ge$5$\sigma$ significance, ($Y$-$J$)$>$0.7 and ($J$-$H$)$>$0.6. This first colour selection will include many late M dwarfs, but it will include nearly all L dwarfs. 137 candidates were identified, all of which also satisfy the following UKIDSS quality criteria: merged class $=$ -1 (i.e. a stellar image profile), ellipticity$<$0.3 in both J band images, and pperrbits$<$256 in both J band images. We then used the optical data from SDSS DR8 to refine the selection. \citet{hawley02} presented the average ($r$-$i$), ($i$-$z$), ($z$-$J$) and ($i$-$J$) colours for objects of spectral types M0 to T6 that were identified in SDSS. The plots shown in figure 8 of that work demonstrate that the ($i$-$J$) colour is best for spectrophotometric typing, and in particular for distinguishing M dwarfs from L dwarfs, so we used this colour to define our final sample. Their $J$ band data were presented on the 2MASS system, so we put them on the MKO system before making a colour cut, which was ($i$-$J$)$\ge$4.4. Only 21 objects remained, all of which passed visual inspection for defects and blending. All of these 21 also have $i$-$z$$>$1.8 and they are drop outs in the $u$ and $g$ passbands, which is consistent with L dwarf status. They are listed in Table \ref{Ltable}.

\begin{table*}
\centering
\small
\begin{tabular}{|l|c|c|c|c|c|c|c|c|c|}
\hline
  \multicolumn{1}{|c|}{$Name$} &
  \multicolumn{1}{|c|}{$\mu_{\alpha} \cos \delta$} &
  \multicolumn{1}{|c|}{$\mu_{\delta}$} &
  \multicolumn{1}{c|}{J} &
  \multicolumn{1}{c|}{J-H} &
  \multicolumn{1}{c|}{H-K} &
  \multicolumn{1}{c|}{i-J} &
  \multicolumn{1}{c|}{i-z} &
  \multicolumn{1}{c|}{Estimated} &
  \multicolumn{1}{c|}{Actual} \\
   & & & & & & & & type & type\\
\hline
  ULAS J073933.51+230709.4 &  $ -84 \pm 14 $  &  $ -122 \pm 14 $  & 18.143 & 0.82 & 0.68 & 6.14 & 3.3 & $>$L7 & \\
  ULAS J075656.40+231456.6 &  $ 162 \pm 10 $  &  $ -154 \pm 10 $  & 16.958 & 1.14 & 0.83 & 5.95 & 3.17 & $>$L7 & L3.5$^a$\\
  ULAS J080441.08+182611.8 &  $ -145 \pm 9 $  &  $ -65 \pm 7 $  & 17.545 & 1.03 & 0.71 & 5.72 & 3.26 & $>$L6 & \\
  ULAS J083023.24+235538.6 &  $ 89 \pm 7 $  &  $ -123 \pm 7 $  & 17.418 & 1.16 & 0.77 & 5.64 & 3.09 & $>$L6 & \\
  ULAS J092933.50+342952.1 &  $ -231 \pm 13 $  &  $ -80 \pm 11 $  & 16.743 & 1.08 & 0.88 & 5.64 & 2.74 & L6 & L8$^b$\\
  ULAS J093336.29+333701.9 &  $ -23 \pm 9 $  &  $ -221 \pm 10 $  & 17.154 & 0.66 & 0.72 & 4.46 & 1.97 & L0 & \\
  ULAS J145949.58+330125.1 &  $ 83 \pm 9 $  &  $ -95 \pm 10 $  & 16.702 & 0.61 & 0.55 & 4.57 & 2.16 & L1 & \\
  ULAS J150231.71+312056.5 &  $ -10 \pm 11 $  &  $ -88 \pm 10 $  & 17.911 & 0.84 & 0.61 & 4.48 & 2.38 & L0 & \\
  ULAS J152225.03+304917.2 &  $ -35 \pm 10 $  &  $ -61 \pm 9 $  & 18.043 & 0.77 & 0.67 & 4.58 & 2.74 & L1 & \\
  ULAS J153158.93+282954.7 &  $ -79 \pm 12 $  &  $ 25 \pm 12 $  & 18.517 & 0.63 & 0.02 & 4.44 & 2.29 & $>$L0 & \\
  ULAS J154432.76+265551.6 &  $ -87 \pm 11 $  &  $ 95 \pm 15 $  & 16.223 & 0.92 & 0.78 & 5.09 & 2.22 & L3 & \\
  ULAS J161626.46+221859.4 &  $ -51 \pm 7 $  &  $ 25 \pm 6 $  & 17.462 & 1.06 & 0.77 & 5.58 & 2.87 & $>$L6 & L5$^a$\\
  ULAS J162052.30+275115.7 &  $ 10 \pm 12 $  &  $ -145 \pm 10 $  & 17.609 & 0.85 & 0.68 & 4.85 & 2.27 & L2 & \\
  ULAS J162339.03+253511.3 &  $ -152 \pm 6 $  &  $ 1 \pm 5 $  & 17.121 & 1.12 & 0.84 & 5.12 & 1.95 & L3 & \\
  ULAS J163352.78+305223.1 &  $ -25 \pm 9 $  &  $ -113 \pm 8 $  & 16.626 & 0.74 & 0.62 & 4.53 & 1.96 & L1 & L2$^c$\\
  ULAS J163713.53+303808.4 &  $ -142 \pm 9 $  &  $ 54 \pm 6 $  & 17.375 & 0.71 & 0.51 & 4.90 & 1.93 & L3 & \\
  ULAS J163836.80+281003.0 &  $ 18 \pm 6 $  &  $ -111 \pm 7 $  & 17.067 & 0.65 & 0.58 & 4.45 & 1.82 & L0 & \\
  ULAS J164131.57+282015.8 &  $ -30 \pm 5 $  &  $ 64 \pm 8 $  & 17.018 & 0.93 & 0.8 & 4.58 & 1.7 & L1 & \\
  ULAS J164301.34+322407.2 &  $ -54 \pm 7 $  &  $ 121 \pm 10 $  & 16.385 & 0.61 & 0.48 & 4.62 & 2.08 & L1 & \\
  ULAS J164456.00+311228.8 &  $ -6 \pm 7 $  &  $ -133 \pm 6 $  & 16.538 & 0.65 & 0.55 & 4.41 & 2.08 & L0 & \\
  ULAS J164522.04+300406.8 &  $ -74 \pm 3 $  &  $ -67 \pm 8 $  & 15.08 & 0.85 & 0.71 & 4.69 & 1.81 & L2 & L3$^d$\\
\hline
\multicolumn{10}{|l|}{$^a$ \citet{knapp04}}\\
\multicolumn{10}{|l|}{$^b$ \citet{kirkpatrick00}}\\
\multicolumn{10}{|l|}{$^c$ \citet{zhang10}}\\
\multicolumn{10}{|l|}{$^d$ \citet{cruz07}}\\
\end{tabular}
\caption{\textbf{L Dwarfs in DR8} Proper motion values are in units of $mas~yr^{-1}$. Near infrared photometry is UKIDSS LAS DR8, J band is first epoch. Optical photometry is SDSS DR8.}
\label{Ltable}
\normalsize
\end{table*}

\begin{figure*}
\begin{center}
\begin{tabular}{c}
\epsfig{file=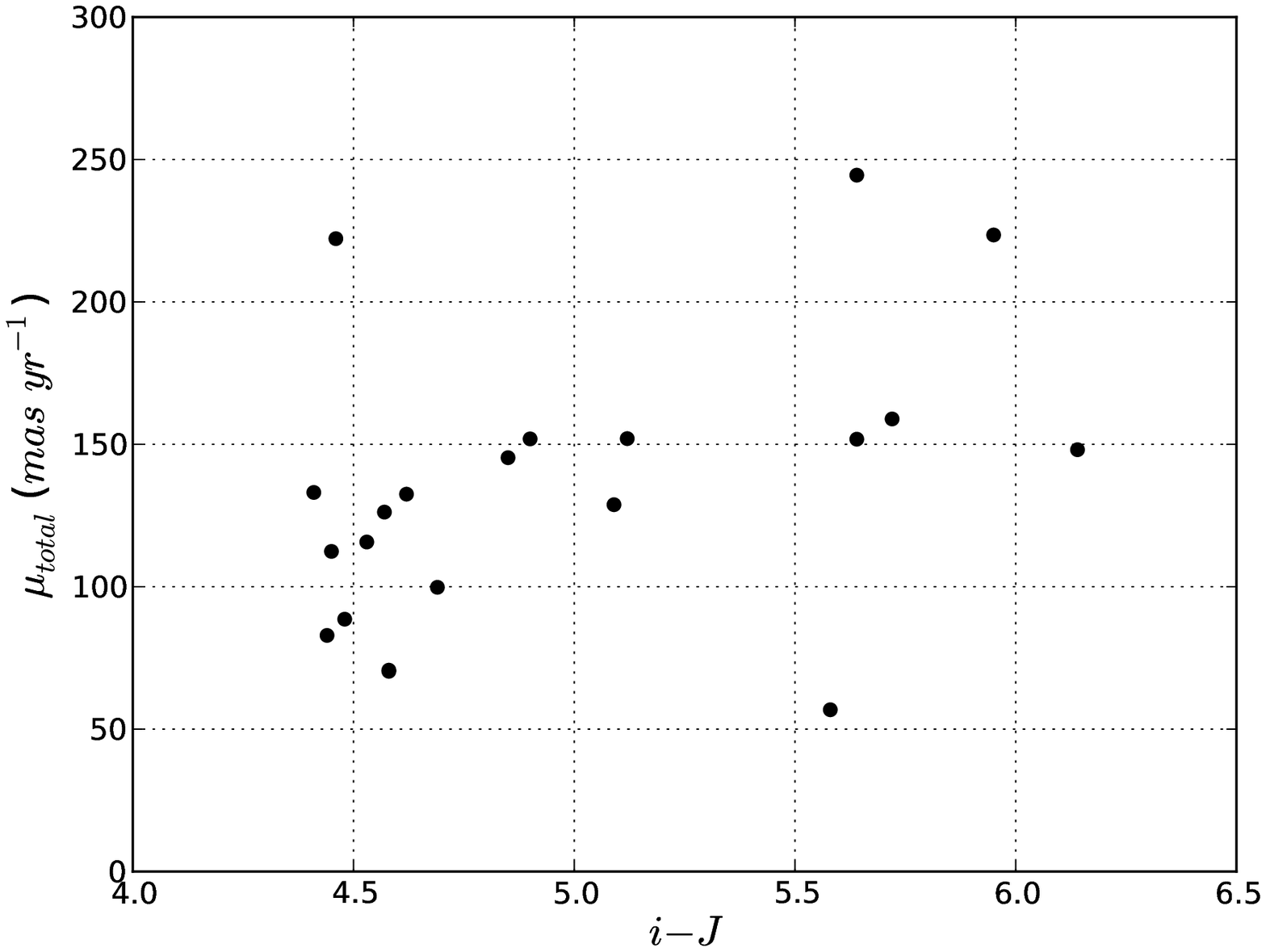,width=0.5\linewidth,clip=}
\epsfig{file=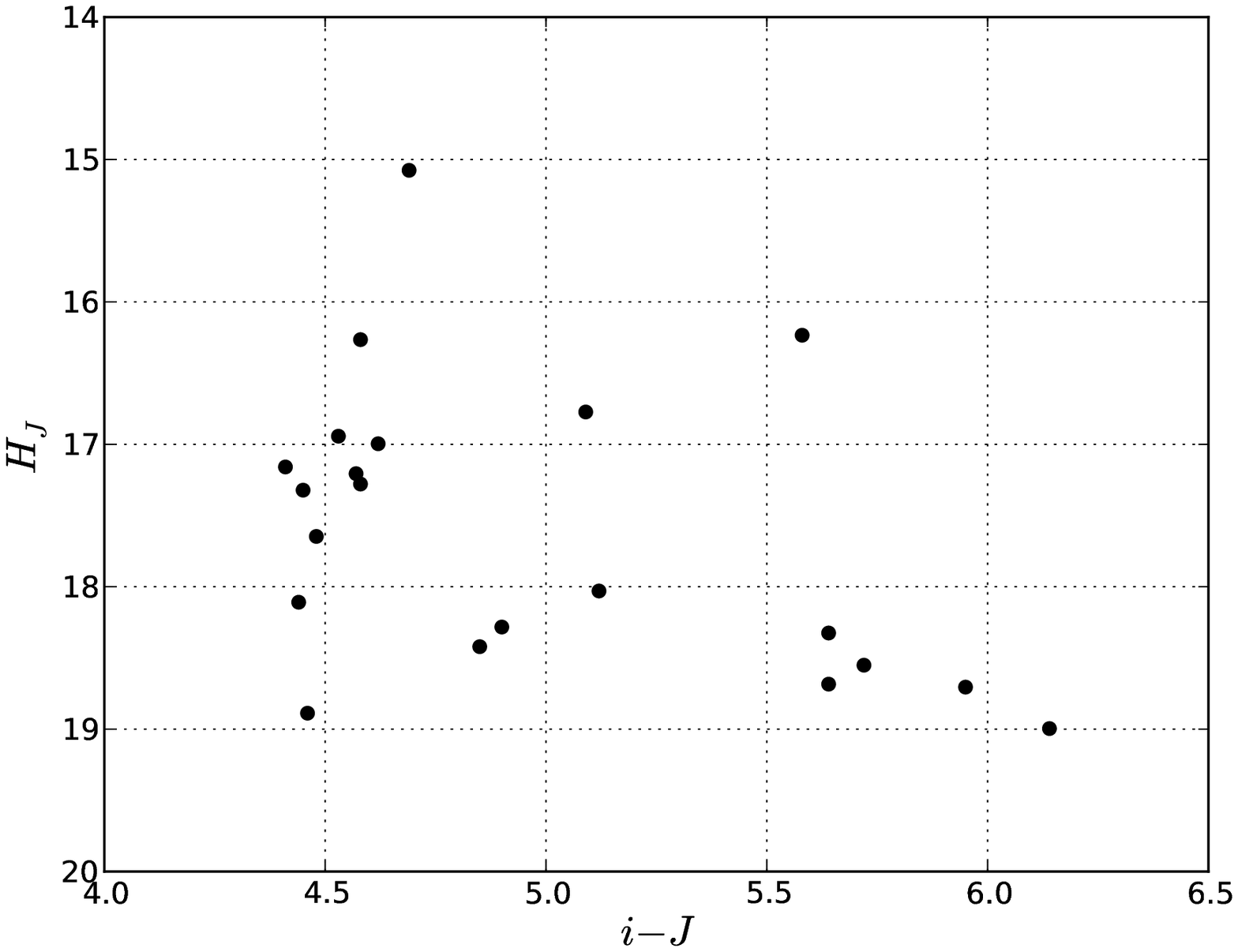,width=0.5\linewidth,clip=}
\end{tabular}
\caption{$i-J$ colour vs. total proper motion ({\it left}) and J band reduced proper motion ({\it right}) for the 21 L dwarfs presented in Table \ref{Ltable}.}
\label{philfigs}
\end{center}
\end{figure*}

In Figure \ref{philfigs} we plot the proper motion against ($i$-$J$) colour (left panel) and the reduced proper motion, $H_J$, against ($i$-$J$) colour. There is a trend for the reddest objects to have the largest reduced proper motions as would be expected if the reddest objects tend to be the coolest and least luminous.

A search of the SIMBAD database shows that 6 of these 21 L dwarf candidates are known in the literature: 5 are known L dwarfs and one, ULAS J154432.76+265551.6, is an L dwarf photometric candidate previously identified using SDSS and UKIDSS photometry \citep{zhang09}. The absence of late M dwarfs indicates that our selection has been successful, despite the significant scatter in the colours of late M and L dwarfs, and the larger volume probed for earlier types in a magnitude limited sample (limited by the sensitivity of UKIDSS and SDSS in this case). Of the 5 known L dwarfs in Table \ref{Ltable}, ULAS J092933.50+342952.1 is type L8 \citep{kirkpatrick00}, ULAS J163352.78+305223.1 is type L2 \citep{zhang10}, ULAS J164522.04+300406.8 is type L3 \citep{cruz07}, ULAS J075656.40+231456.6 is type L3.5 and ULAS J161626.46+221859.4 is type L5 (both from \citealp{knapp04}). A comparison between these spectral types and spectrophotometric types that we inferred from the ($i$-$J$) colour showed agreement to within 1 or 2 subtypes in 4 out of 5 cases, so we list spectrophotometric types for the 21 L dwarf candidates in the table, quoting only limits for objects with photometric uncertainty $>$0.3 mag in the $i$ magnitude.

\subsection{T dwarfs}
\citet{burningham13} present proper motions from our catalogue for 128 UKIDSS T dwarfs, including two new benchmark T dwarfs; LHS 6176B and HD 118865B. 

We also investigated the characteristic population age of late T dwarfs \citep{smith13} in response to current atmospheric models suggesting they are young and low mass (\citealt{leggett09}, \citeyear{leggett10}, \citeyear{leggett12}). For this we used tangential velocity data calculated using proper motions from this and other catalogues and spectrophotometric distances where parallax data were unavailable (\citealt{marocco10}, \citealt{andrei11}, \citealt{dupuy12}, and \citealt{kirkpatrick12}). We concluded that the kinematic age of the population is older than that predicted by the models and, ultimately, more benchmarks are needed to anchor them.

\citet{pinfield12} presented a proper motion for the T8.5 dwarf WISEP J075003.84$+$272544.8 from a prototype version of our catalogue. We note that WISEP J075003.84$+$272544.8 was independently identified by us in a search for T dwarfs with high reduced proper motions shortly before its publication by the WISE team. We now provide the most up to date proper motion from this final revision of our catalogue of $-736\pm13~mas~yr^{-1}$ and $-195\pm15~mas~yr^{-1}$ in $\mu_{\alpha} \cos \delta$ and $\mu_{\delta}$ respectively (the uncertainty is reduced and the proper motion difference is within the uncertainties).

Comparison of our new measured proper motion for the halo T dwarf candidate ULAS J1319+1209 with that reported in \citet{murray11} highlights a significant discrepancy. We measure a considerably lower proper motion, which suggests kinematics most consistent with membership of the Galactic disk, rather than the halo. The previous measurement by \citet{murray11} appears to be in error, with the likely source a dramatic underestimate of the uncertainty in the centroid in its second epoch imaging combined with a relatively short epoch baseline between the observations. Our new measurement benefits from considerably higher precision thanks to deeper UKIDSS second epoch imaging, and a $\sim$5 year epoch baseline, compared to 0.8 years in the \citet{murray11} case.

Results of searches for new T dwarfs with large proper motions, including the identification of two candidate thick disk/halo members, will be reported in a future publication.

\subsection{Brown Dwarf Benchmark Searches}

We undertook various searches for benchmark ultracool dwarfs, the parameters of which and known objects recovered are shown here. Given that parameters derived from atmospheric models of ultracool dwarfs are uncertain it is helpful to constrain them through associations with objects in a common system \citep{pinfield06}. See Section \ref{newbenchmarkcands} for results.

\subsubsection{L Dwarfs in DR8}

We performed a search for binary companions to the 21 L dwarf candidates presented in Section \ref{ldwarfs} by cross matching against the optical proper motion catalogue of \citet{munn04}, which has higher precision than the UKIDSS LAS catalogue, owing to the much longer time baseline. A cross match radius of 1000" was used, which is sufficient to detect candidates within 20,000 AU in every case, even if a `near' distance corresponding to spectral type L9 is assumed for each candidate. We found 61 sources as possible companions to 3 sources from Table \ref{Ltable} that have proper motion vectors consistent to within 2$\sigma$. (This was defined by computing the difference between the L dwarf's UKIDSS proper motion vector and the companion star's Munn catalogue proper motion vector and determining whether the length of the resulting vector is consistent with zero, to within 2$\sigma$). Of these 61 possible companions, 22 are listed as non-stellar sources in UKIDSS (i.e. the parameter {\it pstar $<$0.5}) usually indicating an extended source. We estimated spectrophotometric types and distances to the remaining 39 sources using their SDSS fluxes and the data of \citet{finlator00} and \citet{west11} and found that every candidate was ruled out either because its heliocentric distance was far too great to be a companion to the L dwarf, or because the projected separation at the distance of the star was $>$50,000 AU.

We also searched for binary companions via an internal match of the UKIDSS LAS proper motion catalogue, in order to detect any companions that might be too faint to appear in the Munn catalogue. We searched only within a 300" radius, since binaries with cool primaries (M type or later) have not yet been found with separations $>$6000 AU. We using the same criterion that the proper motion vectors agree to within 2$\sigma$ and the companion must have {\it pstar}$>$0.5, and an additional criterion that the candidate companion must have a brighter $J$ magnitude than the L dwarf. We found 6 candidates, only 1 of which (a star with mid M-type colours) had not been detected and ruled out in the previous match against the Munn catalogue. This object is too distant to be a companion to the L dwarf. 
We therefore conclude that none of the 21 L dwarf candidates have stellar companions.

\subsubsection{L and Early T Dwarfs in the Full Survey}\label{bdbinarysearches}

We searched the full 1500 deg$^2$ of data for new L and T candidates, in particular new benchmark objects, we selected sources with $Y-J~>~0.8$ that were also classified as stellar in both J band images and fell into one or more of the following groups:
\begin{description}
\item No H band detection, stellar merged class, and proper motion $>$ 350 $mas~yr^{-1}$ (21 sources)
\item $J-H~>~0.8$, and proper motion $>$ 500 $mas~yr^{-1}$ (6 sources)
\item $J-K~>~1.4$, and proper motion $>$ 500 $mas~yr^{-1}$ (17 sources)
\item $J-H~>~0.6$, $H-K~>~0.5$ and proper motion $>$ 500 $mas~yr^{-1}$ (12 sources)
\end{description}
There were a further 21 sources with $Y-J~>~0.8$ and proper motion $>$ 500 $mas~yr^{-1}$ that did not meet the other near infrared colour cuts, we elected to take a closer look at these as well since there were relatively few and sources with such large proper motions are often interesting in some way. These groups, taking into account sources in multiple groups, make 57 sources for further study.

We matched to the UKIDSS late T dwarf catalogue maintained by Ben Burningham and we find 9 matches with spectral types ranging from T5 to T9. We also matched to SIMBAD and find a further 7 L dwarfs (including one listed as an L dwarf candidate). This demonstrates the efficacy of our brown dwarf selection criteria.

We searched for benchmark candidates within our sample by retrieving a list of stars with proper motions greater than 350 $mas~yr^{-1}$ from SIMBAD and cross matching with our list of 57 Brown Dwarf candidates within 5' and with proper motion difference tolerances of 50 $mas~yr^{-1}$ independently in RA and Dec.
Four potential benchmark objects have separations ranging from 64" to 111". G 62-33/2MASS J13204427+0409045 are a known K2/L3 binary \citep{faherty10}. Ross 458A/C (\citealt{goldman10}; \citealt{burningham11}) and GJ 576/WISEP J150457.58+053800.1 (\citealt{scholz10}; \citealt{murray11})  are also recovered. The fourth candidate is found to be crosstalk from the bright potential primary upon further inspection of the images. Interestingly the bright star from which our fourth candidate is crosstalk is itself the fainter component of a common proper motion group, the primary being LHS 2968.

\subsubsection{Extended Red Search}\label{blinkingbinaries}

In Section \ref{reliability} we found 834 red (Y-J $>$ 0.7) sources with genuine large motions between the observation epochs ($>$ 500 $mas$). This translates to genuine proper motions down to around 75 $mas~yr^{-1}$. Note that this selection also incorporates most of the objects selected in Section \ref{bdbinarysearches}. Within this sample we identified 33 sources with a possible common proper motion companion within the 1'$\times$1' image.

In an attempt to recover common proper motion companions to these interesting sources we looked within our catalogue for nearby objects (1' radius) with proper motion differences within 1$\sigma$ significance. In practice we found 1$\sigma$ to be sufficient since matches above that were typically of order 5$\sigma$ or greater. We recovered 13 matches meeting these criteria which produced 12 common proper motion pairs since two of the matches were both components of the same pair. We find 4 of these are known common proper motion pairs according to SIMBAD. The 12 pairs are shown in Section \ref{newbenchmarkcands}, Table \ref{intcpmpairs}.

To expand our candidate brown dwarf binary list to include pairs with a primary too bright to be included in our catalogue we used a list of SIMBAD objects with proper motions greater than 100 $mas~yr^{-1}$. We looked for companions to the 834 red sources with genuine large motions within 10' and with $\mu_{\alpha} \cos \delta$ and $\mu_{\delta}$ differences less than 30 $mas~yr^{-1}$ independently of one another. We found 175 matches to these criteria, though we expected many to have been matched to themselves. In order to remove the self matches from our candidate list we rejected those with separation $<$ 5" or J mag difference $<$ 0.5, which should leave only genuine pairs, variable sources, or those with very high proper motion. Thirty one candidate pairs survived this cut. Of these we found 5 pairs had been identified in our internal search and 2 high proper motion single stars had survived the previous cut, leaving 24 systems (see Section \ref{newbenchmarkcands}, Table \ref{simbadcpmpairs}).

\section{New Candidate Benchmarks}\label{newbenchmarkcands}

Amongst the 36 sources in Tables \ref{intcpmpairs} and \ref{simbadcpmpairs} there are 29 ultracool benchmark binary candidates, of which 15 are new and survive a test for common distance. Below we discuss those for which the primary has a distance in the literature and rule out 3 candidates. All of the remainder have spectrophotometric distances consistent with binarity (see Tables \ref{internaldistances} and \ref{simbaddistances}). Note that unless stated otherwise Y, J, H, and K magnitudes in this section are on the MKO system and J band photometry is UKIDSS first epoch.

\subsubsection*{G 151-59}
G 151-59 has a Hipparcos parallax of 12.63$\pm$2.21 $mas$, placing it at a distance of between 67 and 96 $pc$.  Being relatively bright (2MASS J $=$ 8.98$\pm$0.03) it is a well studied K0 type dwarf with known radial velocity (16.98$\pm$0.20$km~s^{-1}$; \citealt{latham02}) and approximately solar metallicity. If we assume this to be a genuine common proper motion pair then the secondary (ULAS J152557.45-020456.4, J $=$ 17.85$\pm$0.05) must be of type L4 to L6 to place it within the same distance range using the \citet{dupuy12} spectral type to absolute J mag relations. Despite the estimated type of L1 given in Table \ref{simbadcpmpairs} this is not ruled out as a genuine pair given the inherent uncertainty in $i-J$ based spectral types of early to mid L dwarfs. To assess whether G 151-59 and the candidate companion might be a chance projection of two objects at different distances, we loosely followed the method used by \citet{gomes13}. We calculated distances for early L dwarfs (L0-L4) using the LAS J magnitude of our candidate and the spectral type to absolute J mag (MKO) relations of \citet{dupuy12}. We then calculated the expected numbers of such L dwarfs in the volume between $\pm$23\% of each distance and a 46" angular radius using the early L dwarf density of 0.0019 pc$^{-3}$ provided by \citet{cruz03} and the breakdown amongst subtypes provided by fig.12 of that work. The $\pm$23\% distance range is based on the typical spread of 0.3 mag in the absolute J magnitudes (approximately 15\% of the distance) of early L dwarfs added in quadrature to the 17.5\% uncertainty in the distance to G 151-59. We expect to find 4$\times$10$^{-4}$ early L dwarfs. It is therefore clearly improbable that our candidate is present simply due to a chance alignment. When the significance of the proper motion similarity with G 151-59 is also taken into account the chance is lower still. Assuming the candidate is a genuine companion, we calculate a tangential velocity between 75 and 108 $km~s^{-1}$ when the range of possible Hipparcos distances is taken into account. Note that a distance compatible with an L0 dwarf would imply a tangential velocity of order 200 km$~s^{-1}$, which is larger than that of the normal disk population. At the distance range of the potential primary, the pair would have a physical separation of between 3,100 and 4,400 $AU$.

\subsubsection*{10 Vir}
10 Vir (BD+02 2517A) has a Hipparcos parallax of 13.69$\pm$0.31 $mas$ corresponding to a distance of between 71 and 75 $pc$ and an USNO-B I magnitude of 4.7. Assuming this is a genuine common proper motion pair then the secondary must be of type M5 to M8, using the spectral type to MKO absolute J magnitude relation of \citet{dupuy12}, taking into account the uncertainties on the spectral types of those within this range of figure 25 in that work, and its UKIDSS J mag of 14.65$\pm$0.01. This spectral type range is consistent with the estimate given in Table \ref{simbadcpmpairs}, we therefore conclude this to be a likely common proper motion companion. 10 Vir has one known close companion, BD+02 2517B \citep{mason01} though we are unable to recover this object in our catalogue. \citet{mason01} gives a V mag of 13.4 for BD+02 2517B which should be easily visible in the UKIDSS J band image, though we find no source at the given position. This may be explained by the 1909 observation epoch and large proper motion. On inspection of the two epochs of UKIDSS J band images it is apparent that there is a close (4.5" separate) common proper motion companion to 10 Vir. This source is not detected in UKIDSS Y, J and H bands due to the close proximity of 10 Vir, the K band detection (magnitude 12.425$\pm$0.002) may be contaminated by a diffraction spike. If we are to assume that this is BD+02 2517B then we provide an updated position of 12:09:41.73 +01:53:45.28 at the UKIDSS K band epoch of 2008-05-28. We therefore conclude that our late M dwarf common proper motion companion is a likely third, widely separated (10,000 $AU$) component of this system. 

\subsubsection*{HD 115151}
HD 115151 has a Hipparcos parallax of 10.73$\pm$1.16 $mas$ corresponding to a distance of between 84 and 104 $pc$ and a 2MASS J mag of 7.87$\pm$0.03. Assuming this is a genuine common proper motion pair then the secondary must be of type M6 to L1, using the spectral type to MKO absolute J magnitude relation of \citet{dupuy12}, taking into account the uncertainties on the spectral types of those within this range of figure 25 in that work and its UKIDSS J mag of 15.85$\pm$0.01. This spectral type range is consistent with the estimate given in Table \ref{simbadcpmpairs}, we therefore conclude this to be a likely binary.

\subsubsection*{LP 488-31 \& 2MASS J13272850+0916323}
These binaries are identified in Table 3 of \citet{deacon09} but not commented further upon.

\subsubsection*{BD+13 2724}
The BD+13 2724 binary companion does not have a distance estimate compatible with that of the primary (see Table \ref{simbaddistances}), we therefore rule out these two sources as being part of a common system.

\subsubsection*{SDSS J120331.33-005332.8}
SDSS J120331.33-005332.8 is a G type subdwarf with a heliocentric distance of 378$\pm$35 $pc$ \citep{dierickx10}. This pair have a weak proper motion match and the candidate secondary, a late M dwarf, would be within 100 $pc$ so we have ruled these out as a binary pair.

\subsubsection*{2MASS J13284331+0758378}
2MASS J13284331+0758378 is at first glance a widely separated (10') proper motion match to the M8.5/M6 candidate binary pair in Table \ref{intcpmpairs}. When the \citet{zhang10} distance estimate of 118 $pc$ for the M8.5 dwarf in that system is adopted, the physical separation of that system and 2MASS J13284331+0758378 works out at 70,000 $AU$. 2MASS J13284331+0758378 is likely to be an M7/8 dwarf based on its $i-J$ colour and its distance is therefore incompatible with the M6/M8.6 internal binary pair and we can safely rule it out as a third component.

\begin{landscape}
\begin{table}
\centering
\small
\begin{tabular}{|c|c|c|c|c|c|c|c|c|c|c|l|c|}
\hline
  \multicolumn{1}{|c|}{$\alpha$}   &  
  \multicolumn{1}{c|}{$\delta$}   &  
  \multicolumn{1}{c|}{i}   &  
  \multicolumn{1}{c|}{z}   &  
  \multicolumn{1}{c|}{J}   &  
  \multicolumn{1}{c|}{H}   &  
  \multicolumn{1}{c|}{K}   &  
  \multicolumn{1}{c|}{$\mu_{\alpha} \cos \delta$}   &  
  \multicolumn{1}{c|}{$\mu_{\delta}$}   &  
  \multicolumn{1}{c|}{i-J}   &  
  \multicolumn{1}{c|}{SIMBAD}   &  
  \multicolumn{1}{c|}{Spectral}   &  
  \multicolumn{1}{c|}{Known} \\
  &&&&&&&$mas~yr^{-1}$&$mas~yr^{-1}$&&entry&type&binary\\
\hline
  09:53:24.13   &   +05:27:01.3   &   15.64  $\pm$  0.00   &   14.68  $\pm$  0.00   &   13.13  $\pm$  0.01   &   12.58  $\pm$  0.00   &   12.24  $\pm$  0.00   &   -185  $\pm$  8   &   38  $\pm$  7   &   2.51   &   y   &   M4   &   $^b$\\
  09:53:24.45   &   +05:26:58.7   &   20.20  $\pm$  0.04   &   18.21  $\pm$  0.02   &   15.74  $\pm$  0.01   &   15.05  $\pm$  0.01   &   14.47  $\pm$  0.01   &   -188  $\pm$  9   &   39  $\pm$  6   &   4.46   &   y   &   M9.5   & \\ [0.2cm]
  09:56:14.81   &   +01:44:57.5   &   14.12  $\pm$  0.00   &   13.49  $\pm$  0.00   &   12.11  $\pm$  0.01   &   11.65  $\pm$  0.00   &   11.40  $\pm$  0.00   &   -105  $\pm$  11   &   -182  $\pm$  9   &   2.01   &   y   &   M2   &   $^b$\\
  09:56:13.07   &   +01:45:13.1   &   20.29  $\pm$  0.04   &   18.50  $\pm$  0.04   &   16.35  $\pm$  0.01   &   15.92  $\pm$  0.01   &   15.48  $\pm$  0.02   &   -109  $\pm$  11   &   -177  $\pm$  9   &   3.94   &   y   &   M9   & \\[0.2cm]
  11:58:25.59   &   -01:22:58.9   &   15.34  $\pm$  0.00   &   14.49  $\pm$  0.00   &   12.95  $\pm$  0.01   &   12.49  $\pm$  0.00   &   12.18  $\pm$  0.00   &   -201  $\pm$  8   &   -74  $\pm$  7   &   2.39   &   y   &   \textit{M4}$^a$   & \\
  11:58:24.04   &   -01:22:45.5   &   20.31  $\pm$  0.04   &   18.69  $\pm$  0.04   &   16.64  $\pm$  0.01   &   16.12  $\pm$  0.01   &   15.66  $\pm$  0.02   &   -208  $\pm$  10   &   -75  $\pm$  7   &   3.67   &   y   &   M8   &  \\[0.2cm]
  11:59:48.15   &   +07:06:59.1   &   18.49  $\pm$  0.01   &   17.16  $\pm$  0.01   &   15.30  $\pm$  0.01   &   14.80  $\pm$  0.01   &   14.40  $\pm$  0.01   &   -160  $\pm$  12   &   101  $\pm$  10   &   3.19   &   y   &   M7   &   $^c$\\
  11:59:48.47   &   +07:07:09.1   &   17.81  $\pm$  0.01   &   18.01  $\pm$  0.02   &   17.52  $\pm$  0.04   &   17.58  $\pm$  0.06   &   17.49  $\pm$  0.10   &   -159  $\pm$  12   &   102  $\pm$  10   &   0.29   &   y   &   WD?   & \\[0.2cm]
  12:08:16.83   &   +08:45:27.6   &   17.67  $\pm$  0.01   &   16.08  $\pm$  0.01   &   13.94  $\pm$  0.01   &   13.37  $\pm$  0.00   &   12.90  $\pm$  0.00   &   -122  $\pm$  7   &   -68  $\pm$  9   &   3.73   &   y   &   M9   &  \\
  12:08:15.55   &   +08:45:42.7   &   17.77  $\pm$  0.01   &   17.64  $\pm$  0.01   &   16.75  $\pm$  0.02   &   16.51  $\pm$  0.02   &   16.40  $\pm$  0.03   &   -123  $\pm$  8   &   -66  $\pm$  9   &   1.02   &   y   &   WD   &  \\[0.2cm]
  13:25:13.86   &   +12:30:13.3   &   17.96  $\pm$  0.01   &   16.85  $\pm$  0.01   &   15.11  $\pm$  0.01   &   14.56  $\pm$  0.00   &   14.21  $\pm$  0.01   &   -95  $\pm$  8   &   -41  $\pm$  8   &   2.85   &   n   &   \textit{M5/6}$^a$   &  \\
  13:25:12.44   &   +12:30:22.0   &   20.37  $\pm$  0.04   &   18.70  $\pm$  0.04   &   16.42  $\pm$  0.01   &   15.79  $\pm$  0.01   &   15.31  $\pm$  0.01   &   -98  $\pm$  8   &   -38  $\pm$  8   &   3.95   &   n   &   \textit{M8.5/9}$^a$   &  \\[0.2cm]
  13:28:35.49   &   +08:08:19.5   &   18.34  $\pm$  0.01   &   17.06  $\pm$  0.01   &   15.27  $\pm$  0.01   &   14.74  $\pm$  0.01   &   14.35  $\pm$  0.01   &   -145  $\pm$  8   &   -60  $\pm$  9   &   3.07   &   n   &   \textit{M6}$^a$   &  \\
  13:28:34.69   &   +08:08:18.9   &   20.52  $\pm$  0.04   &   18.62  $\pm$  0.03   &   16.45  $\pm$  0.01   &   15.86  $\pm$  0.01   &   15.33  $\pm$  0.01   &   -145  $\pm$  8   &   -60  $\pm$  9   &   4.07   &   y   &   M8.5   &  \\[0.2cm]
  14:04:40.20   &   -00:40:19.8   &   15.52  $\pm$  0.00   &   14.63  $\pm$  0.00   &   13.06  $\pm$  0.01   &   12.56  $\pm$  0.00   &   12.24  $\pm$  0.00   &   -135  $\pm$  8   &   -79  $\pm$  10   &   2.46   &   n   &   \textit{M4}$^a$   &  \\
  14:04:40.39   &   -00:40:26.9   &   17.17  $\pm$  0.01   &   16.03  $\pm$  0.01   &   14.25  $\pm$  0.01   &   13.72  $\pm$  0.00   &   13.35  $\pm$  0.00   &   -136  $\pm$  8   &   -81  $\pm$  10   &   2.92   &   n   &   \textit{M6}$^a$   &  \\[0.2cm]
  14:20:16.86   &   +12:07:38.9   &   16.18  $\pm$  0.00   &   15.21  $\pm$  0.01   &   13.55  $\pm$  0.01   &   13.09  $\pm$  0.00   &   12.71  $\pm$  0.00   &   -108  $\pm$  7   &   -21  $\pm$  8   &   2.64   &   n   &   \textit{M5}$^a$   &  \\
  14:20:17.83   &   +12:07:53.5   &   20.76  $\pm$  0.07   &   18.86  $\pm$  0.05   &   16.54  $\pm$  0.01   &   15.89  $\pm$  0.01   &   15.35  $\pm$  0.01   &   -113  $\pm$  8   &   -20  $\pm$  8   &   4.22   &   n   &   \textit{L0}$^a$   &  \\[0.2cm]
  14:24:38.98   &   +09:17:09.4   &   20.53  $\pm$  0.06   &   18.46  $\pm$  0.03   &   15.68  $\pm$  0.01   &   14.80  $\pm$  0.00   &   14.07  $\pm$  0.00   &   -221  $\pm$  7   &   -156  $\pm$  4   &   4.85   &   y   &   L4   &   $^d$\\
  14:24:39.04   &   +09:17:13.1   &   14.91  $\pm$  0.01   &   14.89  $\pm$  0.01   &   14.52  $\pm$  0.01   &   14.52  $\pm$  0.00   &   14.58  $\pm$  0.01   &   -211  $\pm$  7   &   -156  $\pm$  4   &   0.39   &   y   &   DA   &  \\[0.2cm]
  14:59:41.64   &   +08:35:07.7   &   18.69  $\pm$  0.01   &   17.63  $\pm$  0.02   &   15.83  $\pm$  0.01   &   15.31  $\pm$  0.01   &   14.98  $\pm$  0.01   &   43  $\pm$  7   &   -85  $\pm$  5   &   2.86   &   n   &   \textit{M5/6}$^a$   &  \\
  14:59:41.92   &   +08:35:13.5   &   20.28  $\pm$  0.04   &   19.13  $\pm$  0.05   &   17.18  $\pm$  0.02   &   16.71  $\pm$  0.04   &   16.37  $\pm$  0.04   &   50  $\pm$  6   &   -85  $\pm$  5   &   3.10   &   n   &   \textit{M6}$^a$   &  \\[0.2cm]
  15:49:51.57   &   +08:57:29.6   &   15.45  $\pm$  0.00   &   14.70  $\pm$  0.00   &   13.20  $\pm$  0.01   &   12.68  $\pm$  0.00   &   12.40  $\pm$  0.00   &   -55  $\pm$  3   &   -100  $\pm$  5   &   2.25   &   n   &   \textit{M4}$^a$   &  \\
  15:49:51.88   &   +08:57:30.7   &   20.03  $\pm$  0.03   &   18.50  $\pm$  0.03   &   16.29  $\pm$  0.01   &   15.76  $\pm$  0.01   &   15.28  $\pm$  0.01   &   -57  $\pm$  4   &   -101  $\pm$  5   &   3.74   &   n   &   \textit{M8}$^a$   &  \\
\hline
\multicolumn{8}{|l|}{$^a$ Indicates an estimated spectral type based on $i-J$ colour using \citet{hawley02} as in Section \ref{ldwarfs}.}\\
\multicolumn{8}{|l|}{$^b$ \citet{zhang10}}\\
\multicolumn{8}{|l|}{$^c$ \citet{deacon09}}\\
\multicolumn{8}{|l|}{$^d$ \citet{becklin88}}\\
\end{tabular}
\caption{\textbf{Candidate Binaries Found Internally} The 12 common proper motion systems identified within the catalogue from section \ref{blinkingbinaries}}
\label{intcpmpairs}
\normalsize
\end{table}
\end{landscape}

\begin{landscape}
\begin{table}
\centering
\small
\begin{tabular}{|l|c|c|c|c|c|c|c|c|c|c|c|}
\hline
  \multicolumn{2}{|l|}{\textit{Primary}} &
  \multicolumn{7}{l|}{\textit{Secondary}} &
  \multicolumn{3}{l|}{\textit{Pair}} \\
  \multicolumn{1}{|l|}{Name}   &  
  \multicolumn{1}{c|}{Spectral}   &  
  \multicolumn{1}{c|}{$\alpha$}   &  
  \multicolumn{1}{c|}{$\delta$}   &  
  \multicolumn{1}{c|}{J}   &  
  \multicolumn{1}{c|}{$\mu_{\alpha} \cos \delta$}   &  
  \multicolumn{1}{c|}{$\mu_{\delta}$}   &  
  \multicolumn{1}{c|}{i-J}   &  
  \multicolumn{1}{c|}{Spectral}   &  
  \multicolumn{1}{c|}{Separation}   &  
  \multicolumn{1}{c|}{$\mu_{diff}$}   &  
  \multicolumn{1}{c|}{Known} \\
  &type &&&& $mas~yr^{-1}$& $mas~yr^{-1}$&&type &"&$\sigma$&binary\\
\hline
  NLTT 21820   					&			&   09:27:53.47   &   +01:49:13.9   &   17.95 $\pm$ 0.03   &   -26 $\pm$ 8   	&   -138 $\pm$ 8   	   &   3.17 $\pm$ 0.10   	&   \textit{M6}$^a$   		&   111   	&   1.7   &   $$	\\
  LP 488-31   					&			&   09:46:12.09   &   +11:16:31.0   &   15.66 $\pm$ 0.01   &   -187 $\pm$ 6   	&   -25 $\pm$ 6   	   &   3.33 $\pm$ 0.01   	&    \textit{M7}$^a$   		&   10   	&   0.5   &  $^b$	\\
  2MASS J10084007+0150537   	&			&   10:09:04.17   &   +01:58:46.5   &   17.99 $\pm$ 0.05   &   34 $\pm$ 7   	&   -163 $\pm$ 10      &    $>$ 3.3   	     	&    $>$ \textit{M6.5}$^a$  &   595   	&   1.5   &   $$	\\
  2MASS J12020964+0742538   	&			&   12:01:59.65   &   +07:35:53.6   &   14.97 $\pm$ 0.01   &   127 $\pm$ 7   	&   -136 $\pm$ 7   	   &   3.69 $\pm$ 0.01   	&   M8   					&   446   	&   1.4   &   $$	\\
  2MASS J12020933+0742477   	&			&   12:01:59.65   &   +07:35:53.6   &   14.97 $\pm$ 0.01   &   127 $\pm$ 7   	&   -136 $\pm$ 7   	   &   3.69 $\pm$ 0.01   	&   M8   					&   438   	&   1.4   &   $$	\\
  SDSS J120331.33-005332.8  	&			&   12:02:53.30   &   -00:56:08.3   &   15.87 $\pm$ 0.01   &   -216 $\pm$ 5   	&   -22 $\pm$ 4   	   &   4.05 $\pm$ 0.03   	&   \textit{M9}$^a$   		&   591   	&   3.3   &   $$	\\
  10 Vir   					&   K3III   &   12:09:49.00   &   +01:52:38.3   &   14.65 $\pm$ 0.01   &   57 $\pm$ 8   	&   -192 $\pm$ 4   	   &   3.32 $\pm$ 0.01   	&    \textit{M7}$^a$  		&   137   	&   2.2   &   $$	\\
  DT Vir   					&   M2Ve	&   13:00:41.73   &   +12:21:14.7   &   16.68 $\pm$ 0.01   &   -639 $\pm$ 9   	&   -24 $\pm$ 10   	   &   6.60 $\pm$ 0.57   	&   T8.5   					&   105		&   1.8   &   $^c$	\\
  HD 115151   					&   G5 		&   13:15:13.10   &   +10:41:57.6   &   15.85 $\pm$ 0.01   &   -154 $\pm$ 7   	&   -20 $\pm$ 6   	   &   3.64 $\pm$ 0.02   	&    \textit{M8}$^a$   		&   39   	&   0.7   &   $$	\\
  G 63-23   					&   K5  	&   13:20:41.49   &   +09:57:49.7   &   13.65 $\pm$ 0.01   &   -251 $\pm$ 8   	&   -142 $\pm$ 6   	   &   3.81 $\pm$ 0.01   	&   M8   					&   169   	&   0.1   &   $^d$	\\
  LHS 2722   					&   K2V  	&   13:20:43.97   &   +04:09:06.4   &   15.18 $\pm$ 0.01   &   -498 $\pm$ 9   	&   199 $\pm$ 10   	   &   4.71 $\pm$ 0.03   	&   L3   					&   68  	&   0.8   &   $^d$	\\
  2MASS J13272850+0916323   	&			&   13:27:26.77   &   +09:16:05.6   &   14.52 $\pm$ 0.01   &   -143 $\pm$ 5   	&   -67 $\pm$ 4   	   &   3.43 $\pm$ 0.01   	&    \textit{M7}$^a$   		&   37   	&   0.3   &  $^b$	\\
  2MASS J13284331+0758378   	&			&   13:28:34.69   &   +08:08:18.9   &   16.45 $\pm$ 0.01   &   -145 $\pm$ 8   	&   -60 $\pm$ 9   	   &   4.07 $\pm$ 0.04   	&   M8.5   					&   595   	&   1.3   &   $$	\\
  2MASS J13284331+0758378   	&			&   13:28:35.49   &   +08:08:19.5   &   15.27 $\pm$ 0.01   &   -145 $\pm$ 8   	&   -60 $\pm$ 9   	   &   3.07 $\pm$ 0.01   	&    \textit{M6}$^a$   		&   593   	&   1.3   &   $$	\\
  BD+13 2724   					&   G5  	&   13:54:41.14   &   +12:47:47.5   &   18.29 $\pm$ 0.05   &   -103 $\pm$ 9   	&   10 $\pm$ 11   	   &   3.68 $\pm$ 0.15   	&    \textit{M8}$^a$   		&   574   	&   1.3   &   $$	\\
  LHS 2875   					&   M2.5	&   14:12:11.72   &   -00:35:14.3   &   13.03 $\pm$ 0.01   &   -705 $\pm$ 8   	&   221 $\pm$ 6   	   &   3.63 $\pm$ 0.01   	&   M6   					&   14   	&   1.1   &   $^e$	\\
  2MASS J14493646+0533379   	&			&   14:49:46.20   &   +05:36:53.4   &   17.79 $\pm$ 0.04   &   -107 $\pm$ 10  	&   -135 $\pm$ 10      &   4.35 $\pm$ 0.39   	&   \textit{L1}$^a$   		&   243   	&   1.0   &   $$	\\
  2MASS J14511622+0922464   	&			&   14:51:24.61   &   +09:20:05.0   &   14.35 $\pm$ 0.01   &   -156 $\pm$ 5   	&   -33 $\pm$ 4   	   &   3.28 $\pm$ 0.01   	&    \textit{M6.5}$^a$   	&   204   	&   1.3   &   $$	\\
  G 66-40   					&			&   14:54:08.08   &   +00:53:25.6   &   15.07 $\pm$ 0.01   &   -270 $\pm$ 9   	&   35 $\pm$ 9   	   &   3.84 $\pm$ 0.01   	&    \textit{M8.5}$^a$   	&   23   	&   0.3   &   $$	\\
  2MASS J14552241+0419361   	&			&   14:55:23.27   &   +04:19:48.6   &   16.12 $\pm$ 0.01   &   -137 $\pm$ 11   	&   -39 $\pm$ 10   	   &   3.24 $\pm$ 0.02   	&    \textit{M6.5}$^a$   	&   18   	&   1.4   &   $$	\\
  USNO-B1.0 0988-00251407   	&			&   14:59:35.25   &   +08:57:51.2   &   17.92 $\pm$ 0.03   &   -172 $\pm$ 10   	&   -83 $\pm$ 7   	   &    $>$ 3.4   			&   T4.5 					&   386   	&   1.4   &   $^f$	\\
  LHS 3020   					&   K8V 	&   15:04:57.66   &   +05:38:00.8   &   16.59 $\pm$ 0.02   &   -616 $\pm$ 11   	&   -523 $\pm$ 9   	   &   7.26 $\pm$ 0.51   	&   T6.5   					&   64   	&   1.4   &   $^c$	\\
  G 151-59   					&   K0  	&   15:25:57.45   &   -02:04:56.4   &   17.85 $\pm$ 0.05   &   179 $\pm$ 9   	&   -158 $\pm$ 10      &   4.43 $\pm$ 0.16   	&   \textit{L1}$^a$   		&   46   	&   0.6   &   $$	\\
  TYC 2032-546-1   				&			&   15:32:52.33   &   +28:51:28.5   &   18.39 $\pm$ 0.06   &   -142 $\pm$ 9   	&   -131 $\pm$ 9   	   &    $>$ 2.9   			&    $>$ \textit{M6}$^a$   	&   10   	&   1.5   &   $$	\\
\hline
\multicolumn{8}{|l|}{$^a$ Indicates an estimated spectral type based on $i-J$ colour using \citet{hawley02} as in Section \ref{ldwarfs}.}\\
\multicolumn{8}{|l|}{$^b$ \citet{deacon09}}\\
\multicolumn{8}{|l|}{$^c$ \citet{scholz10}}\\
\multicolumn{8}{|l|}{$^d$ \citet{faherty10}}\\
\multicolumn{8}{|l|}{$^e$ \citet{luyten79a}}\\
\multicolumn{8}{|l|}{$^f$ \citet{dayjones11}}\\
\end{tabular}
\caption{\textbf{Candidate Binaries Found With SIMBAD} The 24 probable common proper motion systems identified in conjunction with SIMBAD from section \ref{blinkingbinaries}}
\label{simbadcpmpairs}
\normalsize
\end{table}
\end{landscape}

\begin{table*}
\centering
\begin{tabular}{|c|c|c|c|c|c|}
\hline
  \multicolumn{1}{|c|}{$\alpha$} &
  \multicolumn{1}{c|}{$\delta$} &
  \multicolumn{1}{c|}{d$_{min}$} &
  \multicolumn{1}{c|}{d$_{max}$} &
  \multicolumn{1}{c|}{d$_{min}$} &
  \multicolumn{1}{c|}{d$_{max}$} \\
  a & a & a $pc$ & a $pc$ & b $pc$ & b $pc$ \\
\hline
  11:58:25.59 & -01:22:58.9 & 50 & 139 & 130 & 130\\
  14:04:40.20 & -00:40:19.8 & 53 & 146 & 53 & 91\\
  15:49:51.57 & +08:57:29.6 & 56 & 155 & 99 & 134\\
  14:20:16.86 & +12:07:38.9 & 55 & 110 & 89 & 111\\
  12:08:16.83 & +08:45:27.6 & 34 & 34 & 41 & 41\\
  13:25:13.86 & +12:30:13.3 & 78 & 226 & 96 & 121\\
  13:28:35.49 & +08:08:19.5 & 84 & 146 & 105 & 105\\
  14:59:41.64 & +08:35:07.7 & 109 & 314 & 203 & 354\\
\hline
\end{tabular}
\caption{\textbf{Internal Binary Distance Estimates} Upper and lower distance estimates for the components of the new binary candidates in Table \ref{intcpmpairs}. Distances are calculated using LAS first epoch J band magnitude and absolute J magnitudes from \citet{dupuy12} ($\geq$M6) and \citet{hawley02} ($<$M6) and the spectral types in Table \ref{intcpmpairs} $\pm$1 subtype where they are $i-J$ based estimates. Note that we find the SDSS and UKIDSS photometry of the secondary component of 1208+0845 fits that of a $\sim$5000 K H-rich white dwarf, we have used this to produce our distance estimate in this case.}
\label{internaldistances}
\end{table*}

\begin{table*}
\centering
\begin{tabular}{|l|c|c|c|c|c|}
\hline
  \multicolumn{1}{|l|}{Name} &
  \multicolumn{1}{c|}{d$_{min}$} &
  \multicolumn{1}{c|}{d$_{max}$} &
  \multicolumn{1}{c|}{$\chi{}^2$} &
  \multicolumn{1}{c|}{d$_{min}$} &
  \multicolumn{1}{c|}{d$_{max}$} \\
  a & a $pc$ & a $pc$ & a & b $pc$ & b $pc$ \\
\hline
  NLTT 21820 & 272 & 817  & 9.95 & 289 & 416\\
  2MASS J14493646+0533379 &  &   & 193.82 & 137 & 179\\
  2MASS J12020964+0742538 &  &   &  & 61 & 61\\
  2MASS J12020933+0742477 &  &   &  & 61 & 61\\
  TYC 2032-546-1 & 119 & 357  & 0.28 & 13 & 508\\
  SDSS J120331.33-005332.8 &  &   &  & 74 & 92\\
  2MASS J10084007+0150537 &  &   & 23.9 & 11 & 323\\
  2MASS J13284331+0758378 &  &   &  & 106 & 106\\
  G 151-59 & 45 & 136  & 0.31 & 118 & 185\\
  G  66-40 & 32 & 96  & 9.25 & 56 & 63\\
  2MASS J13284331+0758378 &  &  &   & 84 & 121\\
  LP  488-31 & 61 & 182  & 1.49 & 83 & 144\\
  2MASS J13272850+0916323 & 38 & 115  & 3.55 & 49 & 86\\
  HD 115151 & 42 & 127  & 0.76 & 81 & 109\\
  BD+13  2724 & 40 & 120  & 0.0 & 248 & 337\\
  2MASS J14511622+0922464 &  &    & 22.18 & 55 & 79\\
  10 Vir &  &  &    & 52 & 91\\
  2MASS J14552241+0419361 & 63 & 189  & 4.27 & 124 & 178\\
\hline
\end{tabular}
\caption{\textbf{Simbad Binary Distance Estimates} Upper and lower distance estimates for the components of the unknown binary candidates in Table \ref{simbadcpmpairs}. Distances to the primaries are calculated using fits of available photometry (B/V, J, H, K) to atmospheric models of main sequence stars. Missing values in the $\chi{}^2$ column indicate that we were unable to fit an $T_{eff}$ to the photometry or the photometry was unavailable. The results of solutions with a $\chi{}^2$ value greater than 10 are not included as we deem them too unreliable. Upper and lower limits of $\pm$50\% are used to take into account photometric scatter and other sources of uncertainty. Note that our distance estimates calculated in this way are consistent with parallax based distances where available. Distance estimates to the secondary components are calculated using the same method as those in Table \ref{internaldistances}. The only binary candidate we are able to rule out with confidence is BD+13  2724.}
\label{simbaddistances}
\end{table*}

\section{Conclusion}\label{conclusion}

We present our UKIDSS LAS derived proper motion catalogue for approximately 1500 deg$^2$ of northern sky. Proper motions range from 0 to our hard proper motion detection limit of 3.3"$~yr^{-1}$ with a typical 1$\sigma$ uncertainties of about 10 $mas~yr^{-1}$ for bright sources (see Figure \ref{pmerrs}). 

We find proper motions to be largely reliable for sources brighter than about magnitude 19 in the J band, with low ellipticity and stellar morphological classification. While the reliability diminishes we are still finding genuine high proper motions at objects fainter than $J=19$. Correlation of proper motions with existing optical catalogues is good, although we note a small percentage (0.5\%) of motions are measured using deprecated frames and their accuracy suffers as a result.

The catalogue has already enabled the identification of a variety of high proper motion sources in particular LSR J0745+2627, WISEP J075003.84$+$272544.8 and two T dwarf benchmarks: LHS 6176B and HD 118865B. In addition, we identify 16 new candidate benchmark ultracool dwarfs which significantly increases the sample of benchmarks.
We are continuing to mine the catalogue for interesting objects and pursuing our own science goals. We would now like to invite the wider community to join us.

\section*{Acknowledgements}
This research was funded in part by the Science and Technology Facilities Council (STFC).
This work is based in part on data obtained as part of the UKIRT Infrared Deep Sky Survey.
The authors would like to acknowledge the Marie Curie 7th European Community Framework Programme grant n.247593 Interpretation and Parameterization of Extremely Red COOL dwarfs (IPERCOOL) International Research Staff Exchange Scheme.
A.H.A. acknowledges CNPq grant PQ306775/2009-3.
This research has made use of the SIMBAD database and VizieR catalogue access tool, operated at CDS, Strasbourg, France.
This research has made use of NASA's Astrophysics Data System Bibliographic Services.
This research has made use of SAOImage \textsc{DS9}, developed by Smithsonian Astrophysical Observatory.

\bibliographystyle{mn2e}
\bibliography{bibliography}

\newpage
\appendix
\section{Catalogue Sample}

\begin{table*}
\begin{tabular}{|r|r|r|r|r|r|r|r|r|r|r|}
\hline
  \multicolumn{1}{|c|}{Line} &
  \multicolumn{1}{c|}{RA} &
  \multicolumn{1}{c|}{Dec} &
  \multicolumn{1}{c|}{Y} &
  \multicolumn{1}{c|}{e\_Y} &
  \multicolumn{1}{c|}{J1} &
  \multicolumn{1}{c|}{e\_J1} &
  \multicolumn{1}{c|}{J2} &
  \multicolumn{1}{c|}{e\_J2} &
  \multicolumn{1}{c|}{H} &
  \multicolumn{1}{c|}{e\_H} \\
\hline
  1 & 0.00606 & -0.673361 & 15.473 & 0.0050 & 14.969 & 0.01 & 14.987 & 0.01 & 14.43 & 0.0050\\
  2 & 0.012379 & -0.605281 & -9.9999949E8 & -9.9999949E8 & 19.708 & 0.219 & 19.429 & 0.179 & 18.809 & 0.195\\
  3 & 0.016784 & -0.799593 & -9.9999949E8 & -9.9999949E8 & 19.2 & 0.142 & 19.294 & 0.159 & -9.9999949E8 & -9.9999949E8\\
  4 & 0.025975 & -0.702338 & 13.481 & 0.0020 & 13.081 & 0.01 & 13.088 & 0.01 & 12.545 & 0.0020\\
  5 & 0.028177 & -1.142162 & 16.55 & 0.0090 & 16.095 & 0.01 & 16.116 & 0.01 & 15.497 & 0.01\\
  6 & 0.062918 & 15.916086 & 14.303 & 0.0030 & 13.892 & 0.01 & 13.785 & 0.01 & 13.156 & 0.0020\\
  7 & 0.070539 & 15.928027 & 15.103 & 0.0040 & 14.555 & 0.01 & 14.591 & 0.01 & 14.074 & 0.0040\\
  8 & 0.102226 & 15.828653 & 16.047 & 0.0070 & 15.565 & 0.014 & 15.561 & 0.01 & 15.067 & 0.0090\\
  9 & 0.108392 & 15.845481 & 19.093 & 0.082 & 18.358 & 0.153 & 18.559 & 0.077 & 18.3 & 0.152\\
  10 & 0.1466 & 15.906866 & 18.214 & 0.038 & 18.086 & 0.122 & 17.719 & 0.037 & 17.216 & 0.057\\
\hline\end{tabular}
\\
\begin{tabular}{|r|r|r|r|r|r|r|r|r|r|r|r|}
\hline
  \multicolumn{1}{|c|}{Line} &
  \multicolumn{1}{c|}{K} &
  \multicolumn{1}{c|}{e\_K} &
  \multicolumn{1}{c|}{J1ell} &
  \multicolumn{1}{c|}{J1PA} &
  \multicolumn{1}{c|}{J2ell} &
  \multicolumn{1}{c|}{J2PA} &
  \multicolumn{1}{c|}{J1class} &
  \multicolumn{1}{c|}{J2class} &
  \multicolumn{1}{c|}{RAPM\_rel} &
  \multicolumn{1}{c|}{DecPM\_rel} &
  \multicolumn{1}{c|}{e\_RAPM\_rel} \\
\hline
  1 & 14.162 & 0.0060 & 0.08 & -72.6 & 0.08 & -26.81 & -1 & -1 & -15.44 & -21.9 & 3.98\\
  2 & -9.9999949E8 & -9.9999949E8 & 0.4 & 63.01 & 0.24 & 61.0 & 1 & -1 & 44.66 & -112.37 & 12.0\\
  3 & -9.9999949E8 & -9.9999949E8 & 0.41 & -17.6 & 0.25 & -88.07 & -7 & -1 & 175.51 & 164.84 & 13.45\\
  4 & 12.329 & 0.0020 & 0.05 & -78.73 & 0.06 & -17.92 & -1 & -1 & -78.6 & -52.84 & 6.89\\
  5 & 15.258 & 0.014 & 0.03 & 151.43 & 0.08 & 153.55 & -1 & -1 & 43.49 & -22.88 & 6.94\\
  6 & 12.906 & 0.0030 & 0.34 & 106.41 & 0.06 & 122.9 & 1 & -1 & -158.3 & 39.32 & 7.94\\
  7 & 13.827 & 0.0050 & 0.02 & 77.86 & 0.04 & 126.85 & -1 & -1 & 81.27 & 32.61 & 7.81\\
  8 & 14.778 & 0.011 & 0.03 & 70.5 & 0.04 & 108.77 & -1 & -1 & -63.07 & -30.77 & 6.96\\
  9 & 17.535 & 0.123 & 0.12 & 127.78 & 0.42 & 124.2 & -1 & 1 & 115.6 & -68.38 & 15.74\\
  10 & 17.106 & 0.084 & 0.33 & 42.7 & 0.06 & 160.08 & -7 & -1 & 67.12 & -55.98 & 11.44\\
\hline\end{tabular}
\\
\begin{tabular}{|r|r|r|r|r|r|r|r|r|l|}
\hline
  \multicolumn{1}{|c|}{Line} &
  \multicolumn{1}{c|}{e\_DecPM\_rel} &
  \multicolumn{1}{c|}{RAPM} &
  \multicolumn{1}{c|}{DecPM} &
  \multicolumn{1}{c|}{e\_RAPM} &
  \multicolumn{1}{c|}{e\_DecPM} &
  \multicolumn{1}{c|}{J1MJDobs} &
  \multicolumn{1}{c|}{EpochBaseline} &
  \multicolumn{1}{c|}{SourceID} &
  \multicolumn{1}{c|}{local} \\
\hline
  1 & 4.89 & -22.33 & -27.04 & 3.98 & 4.89 & 53634.42578 & 6.069462286 & 433867580351 & true\\
  2 & 11.59 & 37.78 & -117.51 & 12.0 & 11.59 & 53634.42578 & 6.069462286 & 433867580667 & false\\
  3 & 13.14 & 168.63 & 159.7 & 13.45 & 13.14 & 53634.42578 & 6.069462286 & 433867580597 & false\\
  4 & 4.06 & -85.49 & -57.98 & 6.89 & 4.06 & 53634.42578 & 6.069462286 & 433867580292 & true\\
  5 & 5.85 & 36.72 & -28.0 & 6.94 & 5.85 & 53634.42578 & 6.069462286 & 433870019429 & true\\
  6 & 7.03 & -163.67 & 31.52 & 7.94 & 7.03 & 54398.37891 & 3.999807474 & 433804633469 & true\\
  7 & 5.89 & 75.9 & 24.81 & 7.81 & 5.89 & 54398.37891 & 3.999807474 & 433804633507 & true\\
  8 & 7.26 & -68.44 & -38.57 & 6.96 & 7.26 & 54398.37891 & 3.999807474 & 433804633203 & true\\
  9 & 15.89 & 110.23 & -76.18 & 15.74 & 15.89 & 54398.37891 & 3.999807474 & 433804633253 & true\\
  10 & 11.87 & 61.75 & -63.78 & 11.44 & 11.87 & 54398.37891 & 3.999807474 & 433804633434 & false\\
\hline\end{tabular}
\caption{\textbf{Catalogue Sample}}\label{sampletable}
\end{table*}

Table \ref{sampletable} gives a sample of ten rows from the catalogue, the column headers correspond to the following:
\begin{description}
  \item \textsl{Line - Links the same line across the splits in this sample table.}
  \item RA - Right Ascension of first epoch J band detection.
  \item Dec - Declination of first epoch J band detection.
  \item Y - UKIDSS DR10 Y magnitude.
  \item e\_Y - Uncertainty on UKIDSS DR10 Y magnitude.
  \item J1 - UKIDSS FITS File first epoch J band magnitude.
  \item e\_J1 - Uncertainty on UKIDSS FITS File first epoch J band magnitude.
  \item J2 - UKIDSS FITS File second epoch J band magnitude.
  \item e\_J2 - Uncertainty on UKIDSS FITS File second epoch J band magnitude.
  \item H - UKIDSS DR10 H magnitude.
  \item e\_H - Uncertainty on UKIDSS DR10 H magnitude.
  \item K - UKIDSS DR10 K magnitude.
  \item e\_K - Uncertainty on UKIDSS DR10 K magnitude.
  \item J1ell - Ellipticity of first epoch J band detection.
  \item J1PA - Position angle of ellipticity of first epoch J band detection.
  \item J2ell - Ellipticity of second epoch J band detection.
  \item J2PA - Position angle of ellipticity of second epoch J band detection.
  \item J1class - Morphological classification of first epoch J band detection.
  \item J2class - Morphological classification of second epoch J band detection.
  \item RAPM\_rel - Relative proper motion in Right Ascension ($\times \cos \delta$).
  \item DecPM\_rel - Relative proper motion in Declination.
  \item e\_RAPM\_rel - Uncertainty on relative proper motion in Right Ascension ($\times \cos \delta$).
  \item e\_DecPM\_rel - Uncertainty on relative proper motion in Declination.
  \item RAPM - Proper motion in Right Ascention ($\times \cos \delta$).
  \item DecPM - Proper motion in Declination.
  \item e\_RAPM - Uncertainty on proper motion in Right Ascention ($\times \cos \delta$).
  \item e\_DecPM - Uncertainty on proper motion in Declination.
  \item J1MJDobs - Modified Julian Date of first epoch observation.
  \item EpochBaseline - Epoch baseline in decimal years.
  \item SourceID - UKIDSS DR10 sourceID (for WSA crossmatching).
  \item local - Local/Global transformation flag (`true' indicates a local transform was used).
\end{description}
Right Ascension, Declination and position angles are in units of decimal degrees. All UKIDSS magnitudes are AperMag3 values (2" aperture), with -9.999E8 as the null value. Proper motions and uncertainties are in units of $mas~yr^{-1}$. Morphological classification flags are as follows:
\begin{description}
  \item 1 - Galaxy
  \item 0 - Noise
  \item -1 - Star
  \item -2 - Probable star
  \item -7 - Bad pixel within 2" aperture
  \item -9 - Saturated.
\end{description}
The CASU standard source extraction documentation contains more information on these morphological classifications.

\end{document}